\documentclass[prd,aps,nofootinbib,tfloatfix,10 pt]{revtex4}
\usepackage{amssymb}
\usepackage{amsmath,graphicx,color,epsfig}
\usepackage{slashed}
\usepackage{multirow}
\usepackage{lineno}
\begin{document}

\title{Imprints of CP violation asymmetries in rare $\Lambda_{b}\to \Lambda\ell^{+}\ell^{-}$ decay in family non-universal $Z^{\prime}$
model} 
\author{M.Ali Paracha$^{1,2}$}
\email{paracha@phys.qau.edu.pk}
\author{Ishtiaq Ahmed$^{1,4}$}
\email{ishitaq@ncp.edu.pk}
\author{M.Jamil Aslam $^{3}$}
\email{jamil@phys.qau.edu.pk} \affiliation{$^{1}$  Laborat\`orio de Física Te\`orica e
Computacional, Universidade Cruzeiro do Sul, 01506-000 S\~ao Paulo,
Brazil}
\affiliation{$^{2}$ Department of Physics, School of Natural Sciences,\\
National University of Science and Technology,Islamabad,Pakistan}
\affiliation{$^{3}$ Department of Physics,\\
Quaid-i-Azam University, Islamabad 45320, Pakistan}
\affiliation{$^{4}$ National Centre for Physics,\\
Quaid-i-Azam University Campus, Islamabad 45320, Pakistan }

\date{\today}
\begin{abstract}
We investigate the exclusive rare baryonic $\Lambda_{b}\to\Lambda\ell^{+}\ell^{-}$ in a family non-universal $Z^{\prime}$ model, which is one of the natural extension of standard model.
Using  transition form factors, calculated in the framework of light cone QCD sum rules, we analyze the effects of polarized and unpolarized CP violation asymmetries for
the said decay. Our results indicate that the value of unpolarized and polarized CP-violation asymmetries are considerable in both $\Lambda_{b}\to\Lambda\mu^{+}\mu^{-}$ and
$\Lambda_{b}\to\Lambda\tau^{+}\tau^{-}$ channels and hence they give clear indication of new physics arising from the neutral $Z^{\prime}$ gauge boson. It is hoped that 
the measurements of these CP-violating asymmetries will not only help us to relate new physics, but also help us to determine the precise values of the parameters of new
gauge boson $Z^{\prime}$.
\end{abstract}

\maketitle

\section{Introduction}
Despite the discovery of the last missing chunk of the Standard Model (SM), the Higgs boson, and its
phenomenological success, there are hints that leads to new physics (NP). The flavour sector is one
of the key area which comes in these paths. Because of the joint efforts at the hadron colliders and at the $B$ factories which
provided us with data of unprecedented precision in this sector, which are not sensitive to small effects in
theoretical calculations that are essential in comparison with the experimental measurements and to see if 
there are any hints of the NP.

There are two approaches which are mainly considered to investigate physics beyond the Standard model.
The first approach is the direct search of new particles, where to produce the particles corresponding to different NP models, the most alluring in this class are different supersymmetric models, the energy of the colliders is raised. The second approach is the indirect search, i.e. to increase the experimental precision on the data of different SM processes where NP effects can manifest themselves. The focus of the two major detectors  ATLAS and CMS experiments at the Large Hadron
Collider (LHC) at CERN is to detect the possible new particles produced at at sufficiently large energy.  However, in the indirect searches, flavour physics plays an important role to investigate physics within and beyond the SM, and the experiment which represents the precision frontier are the LHCb at LHC, the Belle II at the super KEKB and different planned super-B factories will join this arena in future.

In the precision approach, the processes that are suitable to investigate physics within and beyond the SM are the rare decays, particularly the decays which are 
described by the $b\to s(d)$ transitions. The attractive feature of such kind of decays is that they are not allowed at tree level in the SM and possible only at loop 
level  \cite{GIM}. Therefore, these decays serve as an excellent candidates to chalk out the status of new physics beyond the SM. In mesonic sector, rare decays of $B$ mesons has been 
widely studied both theoretically and experimentally in detail \cite{TH,expt}.

It is well known that the predications of the SM results are in good agreement with the current experimental data, however there are still unanswered 
questions in this elegant model, e.g.  CP violation,  hierarchy puzzle, neutrino oscillations, the few to name. To answer these questions a large number of NP models such as extra 
dimension models, different versions of supersymmetric models,  etc exist in literature and the extensive studies 
on the exclusive semileptonic decays of $B$ mesons and $\Lambda_{b}$ baryonic decays both has also been made\cite{AAli,Aliev,AP1,AP2,JAM1,TA1}.

In grand unified theories such as $SU(5)$ or string inspired $E_{6}$ models \cite{GS,GB,EN,JB,VB}, one of the most relevant is the $Z^{\prime}$
scenarios that include the family non-universal $Z^{\prime}$ \cite{EE,PL1} and leptophobic $Z^{\prime}$ models \cite{JLL,BBL}.
Experimental searches of an extra $Z^{\prime}$ boson is an important task of the Tevatron \cite{TTait} and LHC \cite{TGR} experiments.
On the other hand to get the constraints on the $Z^{\prime}$ gauge boson couplings through low energy processes are crucial and
complementary for direct searches $Z^{\prime}\to e^{+}e^{-}$ at Tevatron \cite{AAb}. The most interesting thing about the 
family non universal $Z^{\prime}$ model is the new CP-violating phase which have large effects on 
many FCNC processes \cite{PL1,PL2}, such as $B_{s}-\bar B_{s}$ mixing \cite{ConstrainedZPC1,PL3,UTfit,CB,DL}, and
rare hadronic $B$-meson decays \cite{VB1,VB2,IA}.

In baryonic sector exclusive $\Lambda_{b}\to\Lambda\ell^{+}\ell^{-}$ decays, at quark level are described by $b\to s \ell^+\ell^-$ transition. The main difference between these 
and other mesonic decays are that they can give information about the helicity structure of the effective Hamiltonian for the FCNC $b\to s$ process in the SM and beyond \cite{TM}.
On the experimental side, first observation on rare baryonic $\Lambda_{b}\to\Lambda\mu^{+}\mu^{-}$ decay has been observed by CDF Collaboration \cite{CDF}, and recently this 
decay was also studied by LHCb collaboration \cite{LHCb}. The experimental investigation motivates theoreticians to do more deep analysis of the different physical 
observables such as branching ratio, forward backward asymmetry, single and double lepton polarization asymmetries and $CP$ violation asymmetry in the these decay modes. It will be hoped that such studiers are useful to distingish various extensions of the SM.

 In the present work, we analyze the effects of polarized and unpolarized $CP$ violation asymmetries for  $\Lambda_{b}\to\Lambda\ell^{+}\ell^{-}$ decay in the family non universal $Z^{\prime}$ model developed in \cite{PL1}. From the $CP$ violation asymmetry view point it is important to emphasize here that $b\to s$ transition matrix elements  are proportional to three
 quark coupling matrix elements usually called CKM matrix elements, $V_{tb}V^{\ast}_{ts}$,$V_{cb}V^{\ast}_{cs}$ and $V_{ub}V^{\ast}_{us}$; however due to the unitarity 
 condition, and neglecting the matrix elements  $V_{ub}V^{\ast}_{us}$ in comparision with $V_{tb}V^{\ast}_{ts}$ and $V_{cb}V^{\ast}_{cs}$ the $CP$ asymmetry is highly 
 suppressed in the SM. Therefore, the measurements of $CP$ violating asymmetries in $b\to s$ decays play an important role to find the imprints of the $Z^{\prime}$ model.
 
 The structure of the paper is as follows. In section II we develop a theoretical tool box in which we present the effective Hamiltonian for the decay $b\to s\ell^{+}\ell^{-}$.
 In the same section we present the transition matrix element for the decay $\Lambda_{b}\to\Lambda\ell^{+}\ell^{-}$ decay, and the expressions for unpolarized and 
 polarized $CP$ violation for the said decay in family non-universal $Z^{\prime}$ model. In Section III we discuss the numerical results of the said physical observables.
 The concluding remarks are also presented in the same section.
\section{Theoretical Tool Box}
At quark level the decay $\Lambda_{b}\to\Lambda\ell^{+}\ell^{-}$
($\ell=\mu,\tau$) is governed by the transition $b\to
s\ell^{+}\ell^{-}$, the effective Hamiltonian for such kind of decays at $O(m_{b})$ scale can be written as 
\begin{equation}
H_{eff}=-\frac{4G_{F}}{\sqrt{2}}V_{tb}V^{\ast
}_{ts}{\sum\limits_{i=1}^{10}} C_{i}({\mu })O_{i}({\mu })
\label{effham1}
\end{equation}
 where $G_{F}$ is Fermi coupling constant and $V_{ij}$ are the matrix elements of the CKM matrix. In Eq. (\ref{effham1}), 
 $O_i{(\mu)}$ are the local quark operators and $C_{i}(\mu)$ are the corresponding Wilson coefficients at energy scale $\mu$. The explicit expressions for of the Wilson coefficients at next to leading logrithim order and next to next leading logrithim
 are given in ref \cite{AJB,AJB1,AJB2,FK,MBW,GC,MM,HH,MM1,EL1}. The operators responsible for such kind of decays are $O_{7}$, $O_{9}$ and $O_{10}$ which are summarized in  \cite{AP2}. 
 In terms of the effective Hamiltonian given in Eq. (\ref{effham1}), the quark level amplitude for the said decay in the SM can be written as 
 
 \begin{align}
&\mathcal{M}^{SM}(b \rightarrow s \ell^{+}\ell^{-})
= -\frac{G_{F}\alpha}{\sqrt{2}%
\pi } V_{tb}V_{ts}^{\ast } \bigg\{ C_{9}^{eff}(\bar{s}\gamma _{\mu
}L b)(\bar{\ell} \gamma ^{\mu
}\ell)\notag\\
&+C_{10}^{SM}(\bar{s}\gamma _{\mu }L b)(\bar{\ell}\gamma ^{\mu }\gamma
_{5}\ell)
-2m_{b}C_{7}^{eff}(\bar{s}i\sigma _{\mu \nu }\frac{q^{\nu }}{q^{2}}R b)(\bar{ \ell%
}\gamma ^{\mu }\ell) \bigg\},\label{quark-amplitude}
\end{align}
where $q^{2}$ is the square of the momentum transfer and $\alpha$ is the fine structure constant.

A family non-universal $Z^{\prime }$ boson could be derived
naturally in many extensions of the SM, the most economical way to
get it is to include an additional $U^{\prime }(1)$ gauge symmetry.
This model has been formulated in detail by Langacker and
Pl\"{u}macher \cite{PL1}. In a family non-universal $Z^{\prime }$
model, FCNC transitions $b\rightarrow s\ell ^{+}\ell ^{-}$ could be
induced at tree level because of the non-diagonal chiral coupling
matrix. Assuming that the couplings of right handed
quark flavors with $Z^{\prime }$ boson are diagonal and ignoring $%
Z-Z^{\prime }$ mixing, the effective Hamiltonian of $Z^{\prime}$
part for the decay $b\rightarrow s\ell ^{+}\ell ^{-}$ can be written
as \cite{QCh,KCh1,VB3}
\begin{eqnarray}
\mathcal{H}^{Z^{\prime}}_{eff}(b\to
s\ell^{+}\ell^{-})=-\frac{2G_{F}}{\sqrt{2}}V_{tb}V^{\ast}_{ts}\left[-\frac{B^{L}_{sb}B^{L}_{\ell\ell}}{V_{tb}V^{\ast}_{ts}}(\bar{s}b)_{V-A}
(\bar{\ell}\ell)_{V-A}-\frac{B^{L}_{sb}B^{R}_{\ell\ell}}{V_{tb}V^{\ast}_{ts}}(\bar{s}b)_{V-A}
(\bar{\ell}\ell)_{V+A}\right].\label{1}
\end{eqnarray}
One can also write Eq. (\ref{1}) in the following way
\begin{eqnarray}
\mathcal{H}^{Z^{\prime}}_{eff}(b\to
s\ell^{+}\ell^{-})=-\frac{4G_{F}}{\sqrt{2}}V_{tb}V^{\ast}_{ts}\left[\Lambda_{sb}
C_{9}^{Z^{\prime}}O_{9}+\Lambda_{sb} C_{10}^{Z^{\prime}}O_{10}\right],\label{2}
\end{eqnarray}
where
\begin{eqnarray}
 \Lambda_{sb}=\frac{4\pi e^{-i\phi_{sb}}}{\alpha_{EM}V_{tb}V^{\ast}_{ts}}\label{lamb}\\
 C_{9}^{Z^{\prime}}=|\mathcal{B}_{sb}|S_{LL}|; &&
 C_{10}^{Z^{\prime}}=|\mathcal{B}_{sb}|D_{LL}| \label{C9C10},
\end{eqnarray}
and
\begin{eqnarray}
S^{LR}_{\ell\ell}=B^{L}_{\ell\ell}+B^{R}_{\ell\ell},\notag \\
D^{LR}_{\ell\ell}=B^{L}_{\ell\ell}-B^{R}_{\ell\ell}.\label{4}
\end{eqnarray}
The most economical feature of  the family non-universal $Z^{\prime}$-model
is that operator basis remains the same as in the SM and the only modifications come are in the Wilson coefficients, where $C_{9}$ and $C_{10}$ get modification
while the Wilson coefficient $C_{7}^{eff}$ remains unchanged. 

The total amplitude for the decay $\Lambda_{b}\to\Lambda\ell^{+}\ell^{-}$  is the sum of SM and $Z^{\prime}$ contribution, 
and can be written as follows
\begin{eqnarray}
\mathcal{M}^{tot}(\Lambda_{b}\to\Lambda\ell^{+}\ell^{-})
= -\frac{G_{F}\alpha}{2\sqrt{2}\pi}V_{tb}V_{ts}^{\ast}[\langle\Lambda(k)|\bar s\gamma_{\mu}(1-\gamma_{5})b|\Lambda_{b}(p)\rangle
\{C_{9}^{tot}(\bar\ell\gamma^{\mu}\ell)+C_{10}^{tot}(\bar\ell\gamma^{\mu}\gamma^{5}\ell)\}\notag\\
-\frac{2m_{b}}{q^{2}}C_{7}^{eff}\langle\Lambda(k)|\bar s i\sigma_{\mu\nu}q^{\nu}(1+\gamma^{5})b|\Lambda_{b}(p)\rangle\bar\ell
\gamma^{\mu}\ell],\label{6}
\end{eqnarray}
where $C_{9}^{tot}=C_{9}^{eff}+\Lambda_{sb} C_{9}^{Z^{\prime}}$ and $C_{10}^{tot}=C_{10}^{SM}+\Lambda_{sb} C_{10}^{Z^{\prime}}$

The matrix elements for the decay $\Lambda_{b}\to \Lambda\ell^{+}\ell^{-}$ can be straightforwardly parameterized in terms of the form
factors as follows\cite{Aliev2010}
\begin{eqnarray}
\langle\Lambda(k)|\bar{s}\gamma_{\mu}(1-\gamma_{5})b|\Lambda_{b}(p)\rangle
&=&\bar{u}_{\Lambda(k)}[f_{1}(q^{2})\gamma_{\mu}+i\sigma_{\mu\nu}q^{\nu}f_{2}(q^{2})+q^{\mu}f_{3}(q^{2})-\gamma_{\mu}\gamma_{5}g_{1}(q^{2})\notag\\
&&-i\sigma_{\mu\nu}q^{\nu}\gamma_{5}g_{2}(q^{2})
-q^{\mu}\gamma_{5}g_{3}(q^{2})]u_{\Lambda_{b}(p)},\label{Aliev}\\
\langle\Lambda(k)|\bar{s}i\sigma^{\mu\nu}q^{\nu}(1+\gamma_{5})b|\Lambda_{b}(p)\rangle
&=&\bar{u}_{\Lambda(k)}[f^{T}_{1}(q^{2})\gamma_{\mu}+i\sigma_{\mu\nu}q^{\nu}f^{T}_{2}(q^{2})+q^{\mu}f^{T}_{3}(q^{2})+\gamma_{\mu}\gamma_{5}g^{T}_{1}(q^{2})\notag\\
&&+i\sigma_{\mu\nu}q^{\nu}\gamma_{5}g^{T}_{2}(q^{2})
+q^{\mu}\gamma_{5}g^{T}_{3}(q^{2})]u_{\Lambda_{b}(p)},\label{Aliev1}
\end{eqnarray}
where $f_{i},g_{i}$ and $f^{T}_{i}, g^{T}_{i}$ are the transition form factors for the decay $\Lambda_{b}\to\Lambda$.

By using the matrix elements which are parameterized in terms of transition form factors [Eqs.(\ref{Aliev}) and (\ref{Aliev1})]
  with expression (\ref{6}), the decay amplitude for the decay $\Lambda_{b}\to\Lambda\ell^{+}\ell^{-}$ can be 
  written as
  \begin{eqnarray}
   \mathcal{M}^{tot}(\Lambda_{b}\to\Lambda\ell^{+}\ell^{-})&=&\frac{G_{F}\alpha_{EM}}{2\sqrt{2}\pi}V_{tb}V^{\ast}_{ts}
   \left[\bar{u}_{\Lambda(k)}\left\{\tau^{1}_{\mu}(\bar\ell\gamma^{\mu}\ell)+
   \tau^{2}_{\mu}(\bar\ell\gamma^{\mu}\gamma^{5}\ell)\right\}]u_{\Lambda_{b}(p)}\right]\label{7}
  \end{eqnarray}
 The hadronic functions $\tau^{1}_{\mu}$ and $\tau^{2}_{\mu}$ are given by
 \begin{eqnarray}
  \tau^{1}_{\mu}&=& A(q^{2})\gamma_{\mu} +iB(q^{2})\sigma_{\mu\nu}q^{\nu}+C(q^{2})q_{\mu}-D(q^{2})\gamma_{\mu}\gamma_{5}
  -iE(q^{2})\sigma_{\mu\nu}q^{\nu}\gamma^{5}-F(q^{2})q_{\mu}\gamma^{5}\label{8}\\
  \tau^{2}_{\mu}&=& G(q^{2})\gamma_{\mu} +iH(q^{2})\sigma_{\mu\nu}q^{\nu}+I(q^{2})q_{\mu}-J(q^{2})\gamma_{\mu}\gamma_{5}
  -iK(q^{2})\sigma_{\mu\nu}q^{\nu}\gamma^{5}-L(q^{2})q_{\mu}\gamma^{5}\label{9}
 \end{eqnarray}
 The auxiliary functions from $A(q^{2})$ to $L(q^{2})$ given in Eqs.(\ref{8}) and (\ref{9}) contains both short and long distance 
 effects which are encapsulated in terms of Wilson coefficients and form factors. The explicit form of these functions can
 be written as follows
 \begin{eqnarray}
  A(q^{2})&=& C_{9}^{tot}f_{1}(q^{2})-\frac{2m_{b}}{q^{2}}C_{7}^{eff}f^{T}_{1}(q^{2})\notag\\
  B(q^{2})&=& C_{9}^{tot}f_{2}(q^{2})-\frac{2m_{b}}{q^{2}}C_{7}^{eff}f^{T}_{2}(q^{2})\notag\\
  C(q^{2})&=& C_{9}^{tot}f_{3}(q^{2})-\frac{2m_{b}}{q^{2}}C_{7}^{eff}f^{T}_{3}(q^{2})\notag\\
  D(q^{2})&=& C_{9}^{tot}g_{1}(q^{2})-\frac{2m_{b}}{q^{2}}C_{7}^{eff}g^{T}_{1}(q^{2})\notag\\
  E(q^{2})&=& C_{9}^{tot}g_{2}(q^{2})-\frac{2m_{b}}{q^{2}}C_{7}^{eff}g^{T}_{2}(q^{2})\label{10}\\
  F(q^{2})&=& C_{9}^{tot}g_{3}(q^{2})-\frac{2m_{b}}{q^{2}}C_{7}^{eff}g^{T}_{3}(q^{2})\notag\\
  G(q^{2})&=& C_{10}^{tot}f_{1}(q^{2})\notag\\
  H(q^{2})&=& C_{10}^{tot}f_{2}(q^{2})\notag\\
  I(q^{2})&=& C_{10}^{tot}f_{3}(q^{2})\notag\\
  J(q^{2})&=& C_{10}^{tot}g_{1}(q^{2})\notag\\
  K(q^{2})&=& C_{10}^{tot}g_{2}(q^{2})\notag\\
  L(q^{2})&=& C_{10}^{tot}g_{3}(q^{2})\notag
\end{eqnarray}
 The matrix element for the decay $\Lambda_{b}\to\Lambda\ell^{+}\ell^{-}$  given in Eq. (\ref{7}) is useful to calculate the 
 physical observables. 
  The formula for double differential decay rate can be written as 
  \begin{eqnarray}
   \frac{d^{2}\Gamma(\Lambda_{b}\to\Lambda\ell^{+}\ell^{-})}{d\cos\theta ds}&=& 
   \frac{1}{2M^{3}_{\Lambda}}\frac{2\beta\sqrt{\lambda}}{(8\pi)^3}| \mathcal{M}^{tot}|^{2}\label{10}
  \end{eqnarray}
where $\beta\equiv\sqrt{1-\frac{4m^{2}_{\ell}}{s}}$ and $\lambda=\lambda(M_{\Lambda_{b}},M_{\Lambda},s)\equiv M^{4}_{\Lambda_{b}}
+M^{4}_{\Lambda}+s^{2}-2M^{2}_{\Lambda_{b}}M^{2}_{\Lambda}-2sM^{2}_{\Lambda_{b}}-2sM^{2}_{\Lambda}$. Also $s$ is the square 
of the momentum transfer $q$ and $\theta$ is the angle between lepton and final state baryon in the rest frame of $\Lambda_{b}$.
By using the expression of amplitude given in Eq.(\ref{7}) and integration over $\cos\theta$, one can get the expression of the 
dilepton invariant mass specturm as 
\begin{eqnarray}
 \frac{d\Gamma(\Lambda_{b}\to\Lambda\ell^{+}\ell^{-})}{ds}=\frac{G^{2}_{F}\alpha^{2}_{EM}\beta\sqrt{\lambda}}{2^{13}\pi^{5}M^{3}_{\Lambda_{b}}}
 |V_{tb}V^{\ast}_{ts}|^{2}\mathcal{M}_{1}\label{11}
\end{eqnarray}
with
\begin{eqnarray}
 \mathcal{M}_{1}&=&\{\frac{8}{3}((3\Delta_{1}+(\frac{4m_{\ell}^{2}\lambda}{s}-\lambda))(|A|^{2}+|D|^{2})+12m_{\ell}^{2}%
 (\Delta|A|^{2}+\omega|D|^{2})+3s(\Delta_{1}|B|^{2}+\Delta_{2}|E|^{2})\notag \\
 &&+(12m_{\ell}^{2}\Delta_{3}+\lambda(s-4m_{\ell}^{2}))(|B|^{2}+|E|^{2})+3(\Delta_{1}|G|^{2}+\Delta_{2}|J|^{2})-12m_{\ell}^{2}(\Delta|G|^{2}-\omega|J|^{2})\notag\\
 &&-48m_{\ell}^{2}M_{\Lambda_{b}}M_{\Lambda}(|G|^{2}-|J|^{2})-(\lambda-\frac{4m_{\ell}^{2}\lambda}{s})(|G|^{2}+|J|^{2}) +(s-4m_{\ell}^{2})(|H|^{2}(3\Delta_{1}+\lambda)+|K|^{2}(3\Delta_{2}+\lambda)))\notag\\
&&+32m_{\ell}^{2}s(\omega|I|^{2}+\Delta|L|^{2})-16\Delta(M_{\Lambda_{b}} +M_{\Lambda})(s+2m_{\ell}^{2})(AB^{\ast}+BA^{\ast})-16\omega(s-4m_{\ell}^{2})(M_{\Lambda_{b}}-M_{\Lambda})(JK^{\ast}+KJ^{\ast})\notag\\
 &&-32m_{\ell}^{2}\Delta(M_{\Lambda_{b}}+M_{\Lambda})
 (JL^{\ast}+LJ^{\ast}) +16\omega(s+2m_{\ell}^{2})(M_{\Lambda_{b}}-M_{\Lambda})(DE^{\ast}+ED^{\ast})\}\label{12a}
\end{eqnarray}
where 
\begin{eqnarray*}
 \Delta&\equiv& (M_{\Lambda}-M_{\Lambda_{b}})^{2}-s\\
 \omega&\equiv& (M_{\Lambda}+M_{\Lambda_{b}})^{2}-s\\
 \omega_{1}&\equiv& (M_{\Lambda}+M_{\Lambda_{b}})^{2}+s\\
 \omega_{2}&\equiv& (M_{\Lambda}-M_{\Lambda_{b}})^{2}+s\\
  \omega_{3}&\equiv& (M_{\Lambda_{b}}^{2}-M_{\Lambda}^{2}+6M_{\Lambda_{b}}M_{\Lambda}-s)\\
  \omega_{4}&\equiv& (M_{\Lambda_{b}}^{2}-M_{\Lambda}^{2}-6M_{\Lambda_{b}}M_{\Lambda}-s)\\
 \Delta_{1}&\equiv& (M_{\Lambda}^{2}-M_{\Lambda_{b}}^{2})^{2}-s(4M_{\Lambda_{b}}M_{\Lambda}+s)\\ 
  \Delta_{2}&\equiv& (M_{\Lambda}^{2}-M_{\Lambda_{b}}^{2})^{2}+s(4M_{\Lambda_{b}}M_{\Lambda}-s)\\
  \Delta_{3}&\equiv&  (M_{\Lambda_{b}}^{2}-M_{\Lambda}^{2})^{2}-s(M_{\Lambda_{b}}-M_{\Lambda})^{2}\\
  \Delta_{4}&\equiv&  (M_{\Lambda_{b}}^{2}-M_{\Lambda}^{2})^{2}+s(M_{\Lambda_{b}}-M_{\Lambda})^{2}\\
\end{eqnarray*}

Following the recipe given in ref. \cite{AP2}, one can define the $CP$-violation asymmetry for the decay $\Lambda_{b}\to\Lambda\ell^{+}\ell^{-}$ for both the cases, i.e. with polarized
and unpolarized leptons as
\begin{eqnarray}
 \mathcal{A}_{CP}(\textbf{S}^{\pm}=\textbf{e}^{\pm}_{i})=\frac{\frac{d\Gamma(\textbf{S}^{-})}{ds}-\frac{d\bar\Gamma(\textbf{S}^{+})}{ds}}{\frac{d\Gamma(\textbf{S}^{-})}{ds}+\frac{d\bar\Gamma(\textbf{S}^{+})}{ds}}\label{13}
\end{eqnarray}
where 
\begin{eqnarray*}
 \frac{d\Gamma\textbf({S}^{-})}{ds}&=&\frac{d\Gamma(\Lambda_{b}\to\Lambda\ell^{+}\ell^{-}\textbf({S}^{-}))}{ds}\\
 \frac{d\bar\Gamma\textbf({S}^{+})}{ds}&=&\frac{d\bar\Gamma(\Lambda_{b}\to\Lambda\ell^{+}\ell^{-}\textbf({S}^{+}))}{ds}.\\
\end{eqnarray*}
The analogous expression for $CP$ conjugated differential decay width is given in ref. \cite{AP2}. The expression for $CP$ violation
asymmetry can be obtained by using Eq. (\ref{11}) and Eq. (\ref{12a}), so one gets
\begin{eqnarray}
 \mathcal{A}_{CP}(\textbf{S}^{\pm}=\textbf{e}^{\pm}_{i})&=&\frac{1}{2}\left[\frac{\mathcal{M}_{1}-\mathcal{\bar M}_{1}}{\mathcal{M}_{1}+\mathcal{\bar M}_{1}}\pm\frac{\mathcal{M}^{i}_{1}-\mathcal{\bar M}^{i}_{1}}{\mathcal{M}^{i}_{1}+\mathcal{\bar M}^{i}_{1}}
 \right]\label{14}
\end{eqnarray}
where $i$ represents the longitudinal ($L$), normal ($N$) and transverse ($T$) polarization of the final state leptons.

Also one can write the polarized and unpolarized $CP$ asymmetry as 
\begin{eqnarray}
 \mathcal{A}_{CP}(s)=\frac{\mathcal{M}_{1}-\mathcal{\bar M}_{1}}{\mathcal{M}_{1}+\mathcal{\bar M}_{1}},& &
 \mathcal{A}_{CP}^{i}(s)=\frac{\mathcal{M}_{1}^{i}-\mathcal{\bar M}_{1}^{i}}{\mathcal{M}_{1}^{i}+\mathcal{\bar M}_{1}^{i}}\label{15}
\end{eqnarray}
The normalized $CP$ violation asymmetry can be defined by using the above definition as follows
\begin{eqnarray}
 \mathcal{A}_{CP}(\textbf{S}^{\pm}=e^{\pm}_{i})=\frac{1}{2}[\mathcal{A}_{CP}(s)\pm \mathcal{A}^{i}_{CP}(s)] \label{16a}
\end{eqnarray}
 In Eq.(\ref{16a}) the positive sign in the second term represents to $L$ and $N$ polarizations, and the negative sign is for $T$
 polarization. 
 
 The following are the results of unpolarized $\mathcal{A}_{CP}$ and $\mathcal{A}^{i}_{CP}$ \cite{AP2} 
 \begin{eqnarray}
  \mathcal{A}_{CP}(s)&=&\frac{-2\mathcal{I}m(\Lambda_{sb})\mathcal{Q}(s)}{\mathcal{M}_{1}+2\mathcal{I}m(\Lambda_{sb})\mathcal{Q}(s)}\label{17}\\
  \mathcal{A}^{i}_{CP}(s)&=&\frac{-2\mathcal{I}m(\Lambda_{sb})\mathcal{Q}^{i}(s)}{\mathcal{M}_{1}+2\mathcal{I}m(\Lambda_{sb})\mathcal{Q}^{i}(s)}\label{18}
 \end{eqnarray}
The explicit form of $\mathcal{Q}(s)$ and the $\mathcal{Q}^{i}(s)$ are given below.
\begin{eqnarray}
 \mathcal{Q}(s)&=&\frac{16}{3s}\{\mathcal{H}_{1}\mathcal{I}m(C_{7}C^{Z^{\ast\prime}}_{9})-\mathcal{H}_{2}\mathcal{I}m(C_{9}C^{Z^{\ast\prime}}_{9})
 +\mathcal{H}_{3}\mathcal{I}m(C_{10}C^{Z^{\ast\prime}}_{10})\}\label{19}
\end{eqnarray}
The explicit form of the functions $\mathcal{H}_{1}$,$\mathcal{H}_{2}$ and $\mathcal{H}_{3}$ can be written as 
\begin{eqnarray}
 \mathcal{H}_{1}&=&\{\lambda s(s-4m^{2}_{\ell})(\mathcal{D}_{4}\mathcal{D}^{\ast}_{3}-\mathcal{D}_{10}\mathcal{D}^{\ast}_{9})
 -6s\Delta(M_{\Lambda_{b}}+M_{\Lambda})(s+2m^{2}_{\ell})(\mathcal{D}_{4}\mathcal{D}^{\ast}_{1}+\mathcal{D}_{2}\mathcal{D}^{\ast}_{3})\notag\\
 &&+6s\omega(M_{\Lambda_{b}}-M_{\Lambda})(s+2m^{2}_{\ell})(\mathcal{D}_{10}\mathcal{D}^{\ast}_{7}-\mathcal{D}_{8}\mathcal{D}^{\ast}_{9})\notag \\
&&+3s(s+4m^{2}_{\ell})((\Delta_{1}+\Delta_{3})\mathcal{D}_{4}\mathcal{D}^{\ast}_{3}-(\Delta_{2}+\Delta_{4})\mathcal{D}_{10}\mathcal{D}^{\ast}_{9})\notag\\
&&+3s(\Delta(\omega_{1}+4m^{2}_{\ell})\mathcal{D}_{2}\mathcal{D}^{\ast}_{1}+\omega(\omega_{2}+4m^{2}_{\ell})\mathcal{D}_{8}\mathcal{D}^{\ast}_{7})
 -\lambda(s-4m^{2}_{\ell})(\mathcal{D}_{2}\mathcal{D}^{\ast}_{1}+\mathcal{D}_{8}\mathcal{D}^{\ast}_{7})\}\label{20a}\\
 \mathcal{H}_{2}&=&\{3s(\Delta(\omega_{1}+4m^{2}_{\ell})|\mathcal{D}_{1}|^{2}+\omega(\omega_{2}+4m^{2}_{\ell})|\mathcal{D}_{7}|^{2})
 +\lambda s(s-4m^{2}_{\ell})(|\mathcal{D}_{3}|^{2}+|\mathcal{D}_{9}|^{2})\notag\\
&&-6s\Delta(M_{\Lambda_{b}}+M_{\Lambda})(s+2m^{2}_{\ell})(\mathcal{D}_{1}\mathcal{D}^{\ast}_{3}+\mathcal{D}_{3}\mathcal{D}^{\ast}_{1})
 +3s(s+4m^{2}_{\ell})(\Delta_{1}|\mathcal{D}_{3}|^{2}+\Delta_{4}|\mathcal{D}_{9}|^{2})\notag\\
&&-6s\omega(M_{\Lambda_{b}}-M_{\Lambda})(s+2m^{2}_{\ell})(\mathcal{D}_{7}\mathcal{D}^{\ast}_{9}+\mathcal{D}_{9}\mathcal{D}^{\ast}_{7})\}\label{21a}\\
 \mathcal{H}_{3}&=&\{\lambda s(s-4m^{2}_{\ell})(|\mathcal{D}_{1}|^{2}+|\mathcal{D}_{7}|^{2})+3s(\lambda_{1}|\mathcal{D}_{1}|^{2}
 +\lambda_{2}|\mathcal{D}_{7}|^{2})+6s\Delta(M_{\Lambda_{b}}+M_{\Lambda})(s-4m^{2}_{\ell})(\mathcal{D}_{1}\mathcal{D}^{\ast}_{3}+\mathcal{D}_{3}\mathcal{D}^{\ast}_{1})\notag\\
&&+12sm^{2}_{\ell}\Delta(M_{\Lambda_{b}}+M_{\Lambda})(\mathcal{D}_{7}\mathcal{D}^{\ast}_{11}+\mathcal{D}_{11}\mathcal{D}^{\ast}_{7})
-3s(s-4m^{2}_{\ell})(\Delta_{1}|\mathcal{D}_{3}|^{2}+\Delta_{2}|\mathcal{D}_{9}|^{2})
-12s^{2}m^{2}_{\ell}(\omega|\mathcal{D}_{5}|^{2}+\Delta|\mathcal{D}_{11}|^{2})\notag\\
&&-\lambda s(s-4m^{2}_{\ell})(|\mathcal{D}_{3}|^{2}+|\mathcal{D}_{9}|^{2})
 -6s\omega(M_{\Lambda_{b}}-M_{\Lambda})(s-4m^{2}_{\ell})(\mathcal{D}_{9}\mathcal{D}^{\ast}_{7}+\mathcal{D}_{7}\mathcal{D}^{\ast}_{9})\notag \\
&&+2s\omega m^{2}_{\ell}(M_{\Lambda_{b}}-M_{\Lambda})\mathcal{D}_{1}\mathcal{D}^{\ast}_{5}\}\label{22a}
\end{eqnarray}
 The expressions for polarized $CP$ asymmetry are given below.\\
\subsection {Longitudinal $CP$ violation}
The longitudinal lepton polarization can be written as
\begin{eqnarray}
  \mathcal{A}^{L}_{CP}(s)&=&\frac{-2\mathcal{I}m(\Lambda_{sb})\mathcal{Q}^{L}(s)}{\mathcal{M}_{1}+2\mathcal{I}m(\Lambda_{sb})\mathcal{Q}^{L}(s)}\label{longitudinal}
\end{eqnarray}
with
 \begin{eqnarray}
  \mathcal{Q}^{L}&=&\frac{8}{3}\{\frac{1}{s}(\mathcal{H}^{L}_{1}\mathcal{I}m(C_{7}C^{Z^{\ast\prime}}_{9})-
  \mathcal{H}^{L}_{2}\mathcal{I}m(C_{9}C^{Z^{\ast\prime}}_{9})+\mathcal{H}^{L}_{3}\mathcal{I}m(C_{10}C^{Z^{\ast\prime}}_{10}))+\frac{1}{\sqrt{s}}(\mathcal{H}^{L}_{4}\mathcal{I}m(C_{7}C^{Z^{\ast\prime}}_{10})\notag \\
  &&-\mathcal{H}^{L}_{5}\mathcal{I}m(C_{9}C^{Z^{\ast\prime}}_{10}))\}\label{23}
 \end{eqnarray}
 where $\mathcal{H}^{L}_{1}=\mathcal{H}_{1}$,$\mathcal{H}^{L}_{2}=\mathcal{H}_{2}$ and $\mathcal{H}^{L}_{3}=\mathcal{H}_{3}$. Also $\mathcal{M}_1$ is defined in Eq. (\ref{12a}) and the terms $\mathcal{H}^{L}_{4}$ and $\mathcal{H}^{L}_{5}$ are given as follows
 \begin{eqnarray}
  \mathcal{H}^{L}_{4}&=&\sqrt{s-4m^{2}_{\ell}}\{3s(\Delta_{1}(\mathcal{D}_{4}\mathcal{D}^{\ast}_{3})-\Delta_{2}(\mathcal{D}_{10}\mathcal{D}^{\ast}_{9}))
 +\lambda s(\mathcal{D}_{4}\mathcal{D}^{\ast}_{3}-\mathcal{D}_{10}\mathcal{D}^{\ast}_{9})\notag\\
 &&-6s\Delta(M_{\Lambda_{b}}+M_{\Lambda})(\mathcal{D}_{2}\mathcal{D}^{\ast}_{3}+\mathcal{D}_{4}\mathcal{D}^{\ast}_{1})
  -6s\omega(M_{\Lambda_{b}}-M_{\Lambda})(\mathcal{D}_{8}\mathcal{D}^{\ast}_{9}+\mathcal{D}_{10}\mathcal{D}^{\ast}_{7})+(3\Delta_{2}-\lambda)(\mathcal{D}_{2}\mathcal{D}^{\ast}_{1}+\mathcal{D}_{8}\mathcal{D}^{\ast}_{7})\}\label{24}\\
  \mathcal{H}^{L}_{5}&=& \sqrt{s-4m^{2}_{\ell}}\{((3\Delta_{1}-\lambda)|\mathcal{D}_{1}|^{2}-(3\Delta_{2}+\lambda)|\mathcal{D}_{7}|^{2}
 -6s\Delta(M_{\Lambda_{b}}+M_{\Lambda})(\mathcal{D}_{1}\mathcal{D}^{\ast}_{3}+\mathcal{D}_{3}\mathcal{D}^{\ast}_{1})\notag \\
 &&+6s\omega(M_{\Lambda_{b}}-M_{\Lambda})(\mathcal{D}_{7}\mathcal{D}^{\ast}_{9}+\mathcal{D}_{9}\mathcal{D}^{\ast}_{7})+s((3\Delta_{1}+\lambda)|\mathcal{D}_{3}|^{2}+(3\Delta_{2}+\lambda)|\mathcal{D}_{9}|^{2})\}\label{25}
 \end{eqnarray}
 \subsection{Normal $CP$ violation Asymmetry}
 In case of the normally polarized lepton, the corresponding normal $CP$ violation can be expressed as
 \begin{eqnarray}
  \mathcal{A}^{N}_{CP}(s)&=&\frac{-2\mathcal{I}m(\Lambda_{sb})\mathcal{Q}^{N}(s)}{\mathcal{M}_{1}+2\mathcal{I}m(\Lambda_{sb})\mathcal{Q}^{N}(s)}\label{normal}
\end{eqnarray}
where
 \begin{eqnarray}
\mathcal{Q}^{N}&=&\{\frac{8}{3s}(\mathcal{H}^{N}_{1}\mathcal{I}m(C_{7}C^{Z^{\ast\prime}}_{9})-\mathcal{H}^{N}_{2}\mathcal{I}m(C_{9}C^{Z^{\ast\prime}}_{9})
 +\mathcal{H}^{N}_{3}\mathcal{I}m(C_{10}C^{Z^{\ast\prime}}_{10}))+\frac{4\pi\sqrt{\lambda}}{\sqrt{s}}(\mathcal{H}^{N}_{4}\mathcal{I}m(C_{9}C^{Z^{\ast\prime}}_{10})\notag\\
&&+\mathcal{H}^{N}_{5}\mathcal{I}m(C_{10}C^{Z^{\ast\prime}}_{9})-\mathcal{H}^{N}_{6}\mathcal{I}m(C_{7}C^{Z^{\ast\prime}}_{10}))\}\label{26}
 \end{eqnarray}
and
\begin{eqnarray}
\mathcal{H}^{N}_{1}&=&3\pi\sqrt{\lambda}m_{\ell}s^{3/2}((M_{\Lambda_{b}}-M_{\Lambda})(\mathcal{D}_{10}\mathcal{D}^{\ast}_{1}
-\mathcal{D}_{2}\mathcal{D}^{\ast}_{9})-(\mathcal{D}_{8}\mathcal{D}^{\ast}_{1}-\mathcal{D}_{2}\mathcal{D}^{\ast}_{7}))\notag\\
&&+3\pi\sqrt{\lambda}m_{\ell}s^{3/2}((M_{\Lambda_{b}}^{2}-M_{\Lambda}^{2})(\mathcal{D}_{4}\mathcal{D}^{\ast}_{9}-\mathcal{D}_{10}\mathcal{D}^{\ast}_{3})+(M_{\Lambda_{b}}+M_{\Lambda})(\mathcal{D}_{4}\mathcal{D}^{\ast}_{7}+\mathcal{D}_{8}\mathcal{D}^{\ast}_{3}))\notag\\
&&+6s\Delta(s+2m^{2}_{\ell})(M_{\Lambda_{b}}+M_{\Lambda})(\mathcal{D}_{2}\mathcal{D}^{\ast}_{3}-\mathcal{D}_{4}\mathcal{D}^{\ast}_{1})+\lambda s(s-4m^{2}_{\ell})(\mathcal{D}_{4}\mathcal{D}^{\ast}_{3}-\mathcal{D}_{10}\mathcal{D}^{\ast}_{9}-\mathcal{D}_{2}\mathcal{D}^{\ast}_{1})\notag\\
&&+3s(s+4m^{2}_{\ell})((\Delta_{1}+\Delta_{3})\mathcal{D}_{4}\mathcal{D}^{\ast}_{3}-(\Delta_{2}+\Delta_{4})\mathcal{D}_{10}\mathcal{D}^{\ast}_{9})
 -\lambda(s-4m^{2}_{\ell})\mathcal{D}_{8}\mathcal{D}^{\ast}_{7}\notag\\
&&+6\omega s(s+2m^{2}_{\ell})(M_{\Lambda_{b}}-M_{\Lambda})(\mathcal{D}_{8}\mathcal{D}^{\ast}_{9}-\mathcal{D}_{10}\mathcal{D}^{\ast}_{7})+3s(\Delta(\omega_{1}+4m^{2}_{\ell})\mathcal{D}_{2}\mathcal{D}^{\ast}_{1}+\omega(\omega_{2}+4m^{2}_{\ell})\mathcal{D}_{8}\mathcal{D}^{\ast}_{7})\label{27a}\\
 \mathcal{H}^{N}_{2}&=& \{3\pi\sqrt{\lambda}m_{\ell}s^{3/2}((M_{\Lambda_{b}}+M_{\Lambda})(\mathcal{D}_{7}\mathcal{D}^{\ast}_{3}+\mathcal{D}_{3}\mathcal{D}^{\ast}_{7})-(M_{\Lambda_{b}}-M_{\Lambda})(\mathcal{D}_{1}\mathcal{D}^{\ast}_{9}-\mathcal{D}_{9}\mathcal{D}^{\ast}_{1}))\notag\\
&&-3\pi\sqrt{\lambda}m_{\ell}s^{3/2}(\mathcal{D}_{1}\mathcal{D}^{\ast}_{7}+\mathcal{D}_{7}\mathcal{D}^{\ast}_{1})-6s\Delta(s+2m^{2}_{\ell})(M_{\Lambda_{b}}+M_{\Lambda})(\mathcal{D}_{1}\mathcal{D}^{\ast}_{3}+\mathcal{D}_{3}\mathcal{D}^{\ast}_{1})\notag\\
&&-\lambda(s-4m^{2}_{\ell})(|\mathcal{D}_{1}|^{2}+|\mathcal{D}_{7}|^{2})+6\omega s(s+2m^{2}_{\ell})(M_{\Lambda_{b}}-M_{\Lambda})(\mathcal{D}_{7}\mathcal{D}^{\ast}_{9}+\mathcal{D}_{9}\mathcal{D}^{\ast}_{7})\notag\\
&&+3s(\Delta(\omega_{1}+4m^{2}_{\ell})|\mathcal{D}_{1}|^{2}+\omega(\omega_{2}+4m^{2}_{\ell})|\mathcal{D}_{7}|^{2})+3s(s+4m^{2}_{\ell})((\Delta_{1}+\Delta_{3})|\mathcal{D}_{3}|^{2}\notag\\
&&+(\Delta_{2}+\Delta_{4})|\mathcal{D}_{9}|^{2}+\lambda s(s-4m^{2}_{\ell})(|\mathcal{D}_{3}|^{2}+|\mathcal{D}_{9}|^{2})+3\pi\sqrt{\lambda}m_{\ell}s^{3/2}(M_{\Lambda_{b}}^{2}-M_{\Lambda}^{2})\mathcal{D}_{3}\mathcal{D}^{\ast}_{9}\}\label{28a}\\
 \mathcal {H}^{N}_{4}&=&m_{\ell}\{s^{2}(\mathcal{D}_{3}\mathcal{D}^{\ast}_{5}+\mathcal{D}_{9}\mathcal{D}^{\ast}_{11})-(M_{\Lambda_{b}}^{2}-M_{\Lambda}^{2})(|\mathcal{D}_{1}|^{2}+|\mathcal{D}_{7}|^{2})-s(\mathcal{D}_{1}\mathcal{D}^{\ast}_{5}+\mathcal{D}_{7}\mathcal{D}^{\ast}_{11})\notag\\
&&+s((M_{\Lambda_{b}}-M_{\Lambda})\mathcal{D}_{3}\mathcal{D}^{\ast}_{1}-(M_{\Lambda_{b}}+M_{\Lambda})\mathcal{D}_{9}\mathcal{D}^{\ast}_{7})\}\label{29a}\\
\mathcal{H}^{N}_{5}&=&m_{\ell}\{s(M_{\Lambda_{b}}+M_{\Lambda})(\mathcal{D}_{5}\mathcal{D}^{\ast}_{1}+\mathcal{D}_{7}\mathcal{D}^{\ast}_{9})
+s(M_{\Lambda_{b}}-M_{\Lambda})(\mathcal{D}_{11}\mathcal{D}^{\ast}_{7}+\mathcal{D}_{5}\mathcal{D}^{\ast}_{1})-(M_{\Lambda_{b}}^{2}-M_{\Lambda}^{2})(|\mathcal{D}_{1}|^{2}+|\mathcal{D}_{7}|^{2})\notag \\
&&+s^{2}(\mathcal{D}_{11}\mathcal{D}^{\ast}_{9}+\mathcal{D}_{5}\mathcal{D}^{\ast}_{3})\}\label{29b}\\
 \mathcal{H}^{N}_{6}&=&m_{\ell}\{s(M_{\Lambda_{b}}-M_{\Lambda})(\mathcal{D}_{8}\mathcal{D}^{\ast}_{11}-\mathcal{D}_{4}\mathcal{D}^{\ast}_{1})+s^{2}(\mathcal{D}_{4}\mathcal{D}^{\ast}_{5}-\mathcal{D}_{10}\mathcal{D}^{\ast}_{11})+s(M_{\Lambda_{b}}+M_{\Lambda})(\mathcal{D}_{10}\mathcal{D}^{\ast}_{7}-\mathcal{D}_{2}\mathcal{D}^{\ast}_{5})\notag\\
&&-(M_{\Lambda_{b}}^{2}-M_{\Lambda}^{2})(\mathcal{D}_{2}\mathcal{D}^{\ast}_{1}+\mathcal{D}_{8}\mathcal{D}^{\ast}_{7})\}\label{30a}
\end{eqnarray}
\subsection{Transverse $CP$-Asymmetry}
The transverse $CP$ violation asymmetry can be written as
 \begin{eqnarray}
  \mathcal{A}^{T}_{CP}(s)&=&\frac{-2\mathcal{I}m(\Lambda_{sb})\mathcal{Q}^{T}(s)}{\mathcal{M}_{1}+2\mathcal{I}m(\Lambda_{sb})\mathcal{Q}^{T}(s)}\label{transverse}
\end{eqnarray}
with
\begin{eqnarray}
\mathcal{Q}^{T}(s)&=&\{\frac{8}{3s}(\mathcal{H}^{T}_{1}\mathcal{I}m(C_{7}C^{Z^{\ast\prime}}_{9})-\mathcal{H}^{T}_{2}\mathcal{I}m(C_{9}C^{Z^{\ast\prime}}_{9}))
+4\pi\sqrt{\lambda(s-4m^{2}_{\ell})}(\mathcal{H}^{T}_{3}\mathcal{R}e(C_{9}C^{Z^{\ast\prime}}_{10})\notag \\
&&+\mathcal{H}^{T}_{4}\mathcal{R}e(C_{10}C^{Z^{\ast\prime}}_{9})-\mathcal{H}^{T}_{5}\mathcal{R}e(C_{7}C^{Z^{\ast\prime}}_{10}))
+\frac{4}{3s}(\mathcal{H}^{T^{A}}_{6}\mathcal{I}m(C_{10}C^{Z^{\ast\prime}}_{10})+\mathcal{H}^{T^{B}}_{6}\mathcal{R}e(C_{10}C^{Z^{\ast\prime}}_{10}))\}\label{30b} 
 \end{eqnarray}
 where $\mathcal{H}^{T}_{1}=\mathcal{H}^{L}_{1}$,$\mathcal{H}^{T}_{2}=\mathcal{H}^{L}_{3}$,
 $\mathcal{H}^{T^{A}}_{6}=(\mathcal{H}^{T^{1}}_{6}+\mathcal{H}^{T^{2}}_{6}+\mathcal{H}^{T^{3}}_{6})$ and
 $\mathcal{H}^{T^{B}}_{6}=(\mathcal{H}^{T^{4}}_{6}+\mathcal{H}^{T^{5}}_{6})$.
  The different terms given in above equation can be expressed as
 \begin{eqnarray}.
\mathcal{H}^{T}_{3}&=&m_{\ell}\{(M_{\Lambda_{b}}+M_{\Lambda})(\mathcal{D}_{3}\mathcal{D}^{\ast}_{7}+\mathcal{D}_{7}\mathcal{D}^{\ast}_{3})
+(M_{\Lambda_{b}}^{2}-M_{\Lambda}^{2})(\mathcal{D}_{3}\mathcal{D}^{\ast}_{9}+\mathcal{D}_{9}\mathcal{D}^{\ast}_{3})+s(\mathcal{D}_{3}\mathcal{D}^{\ast}_{9}-\mathcal{D}_{9}\mathcal{D}^{\ast}_{3})\}\label{31}\\
\mathcal{H}^{T}_{4}&=&m_{\ell}\{(M_{\Lambda_{b}}-M_{\Lambda})(\mathcal{D}_{1}\mathcal{D}^{\ast}_{9}+\mathcal{D}_{9}\mathcal{D}^{\ast}_{1})-(M_{\Lambda_{b}}^{2}-M_{\Lambda}^{2})(\mathcal{D}_{3}\mathcal{D}^{\ast}_{9}-\mathcal{D}_{9}\mathcal{D}^{\ast}_{3})\notag\\
&&+s(\mathcal{D}_{3}\mathcal{D}^{\ast}_{9}-\mathcal{D}_{9}\mathcal{D}^{\ast}_{3})-(M_{\Lambda_{b}}+M_{\Lambda})(\mathcal{D}_{3}\mathcal{D}^{\ast}_{7}+\mathcal{D}_{7}\mathcal{D}^{\ast}_{3})\}\label{32}\\
\mathcal{H}^{T}_{5}&=&m_{\ell}\{(M_{\Lambda_{b}}+M_{\Lambda})(\mathcal{D}_{4}\mathcal{D}^{\ast}_{7}+\mathcal{D}_{8}\mathcal{D}^{\ast}_{3})+(M_{\Lambda_{b}}^{2}-M_{\Lambda}^{2}+s)(\mathcal{D}_{4}\mathcal{D}^{\ast}_{9}-\mathcal{D}_{10}\mathcal{D}^{\ast}_{3})\notag\\
&&+(M_{\Lambda_{b}}-M_{\Lambda})(\mathcal{D}_{10}\mathcal{D}^{\ast}_{1}-\mathcal{D}_{2}\mathcal{D}^{\ast}_{1})\}\label{33}\\
\mathcal{H}^{T^{1}}_{6}&=&\{(6s^{4}-24s^{3}m_{\ell}^{2}-24s^{3}M_{\Lambda_{b}}M_{\Lambda}+2s^{2}\Delta_{5})(|\mathcal{D}_{3}|^{2}+|\mathcal{D}_{9}|^{2})
+24s^{3}m_{\ell}^{2}(|\mathcal{D}_{5}|^{2}+|\mathcal{D}_{11}|^{2})\notag\\
&&+12s^{3}((M_{\Lambda_{b}}-M_{\Lambda})\mathcal{D}_{9}\mathcal{D}^{\ast}_{7}-(M_{\Lambda_{b}}+M_{\Lambda})\mathcal{D}_{3}\mathcal{D}^{\ast}_{1})
+48m_{\ell}^{2}M_{\Lambda_{b}}M_{\Lambda}(|\mathcal{D}_{3}|^{2}-|\mathcal{D}_{9}|^{2})\notag\\
&&+12m_{\ell}^{2}((M_{\Lambda_{b}}+M_{\Lambda})(2\mathcal{D}_{3}\mathcal{D}^{\ast}_{1}+\mathcal{D}_{11}\mathcal{D}^{\ast}_{7})
+(M_{\Lambda_{b}}-M_{\Lambda})(2\mathcal{D}_{9}\mathcal{D}^{\ast}_{7}-\mathcal{D}_{5}\mathcal{D}^{\ast}_{1}))\notag\\
&&+12m_{\ell}^{2}(M_{\Lambda_{b}}(M_{\Lambda_{b}}+M_{\Lambda})|\mathcal{D}_{5}|^{2}-(M_{\Lambda_{b}}-M_{\Lambda})^{2}|\mathcal{D}_{11}|^{2})\notag\\
&&+6(M_{\Lambda_{b}}^{2}-M_{\Lambda}^{2})((M_{\Lambda_{b}}+M_{\Lambda})\mathcal{D}_{9}\mathcal{D}^{\ast}_{7}-(M_{\Lambda_{b}}-M_{\Lambda})\mathcal{D}_{3}\mathcal{D}^{\ast}_{1})\}\label{34}\\
\mathcal{H}^{T^{2}}_{6}&=&8m_{\ell}^{2}s\{3(M_{\Lambda_{b}}^{2}-M_{\Lambda}^{2})((M_{\Lambda_{b}}^{2}-M_{\Lambda}^{2})(|\mathcal{D}_{3}|^{2}+|\mathcal{D}_{9}|^{2})
+(M_{\Lambda_{b}}+M_{\Lambda})(\mathcal{D}_{5}\mathcal{D}^{\ast}_{1}-2\mathcal{D}_{9}\mathcal{D}^{\ast}_{7})\notag\\
&&+(M_{\Lambda_{b}}-M_{\Lambda})(2\mathcal{D}_{3}\mathcal{D}^{\ast}_{1}-\mathcal{D}_{11}\mathcal{D}^{\ast}_{7}))
+\lambda(|\mathcal{D}_{3}|^{2}+|\mathcal{D}_{9}|^{2})\}\label{35}\\
\mathcal{H}^{T^{3}}_{6}&=&\{(2s\lambda-8m^{2}_{\ell}\lambda)(|\mathcal{D}_{1}|^{2}+|\mathcal{D}_{7}|^{2})+(24sm^{2}_{\ell}-6s)(\Delta_{1}|\mathcal{D}_{1}|^{2}-\Delta_{2}|\mathcal{D}_{7}|^{2})\notag\\
&&+24sm^{2}_{\ell}(\omega_{4}|\mathcal{D}_{1}|^{2}+\omega_{3}|\mathcal{D}_{7}|^{2})+12s(s-4m^{2}_{\ell})(\Delta(M_{\Lambda_{b}}+M_{\Lambda})\mathcal{D}_{1}\mathcal{D}^{\ast}_{3}-\omega(M_{\Lambda_{b}}-M_{\Lambda})\mathcal{D}_{7}\mathcal{D}^{\ast}_{9})\notag\\
&&+24sm^{2}_{\ell}(\Delta(M_{\Lambda_{b}}+M_{\Lambda})\mathcal{D}_{7}\mathcal{D}^{\ast}_{11}+\omega(M_{\Lambda_{b}}-M_{\Lambda})\mathcal{D}_{1}\mathcal{D}^{\ast}_{5})\}\label{36}\\
\mathcal{H}^{T^{4}}_{6}&=& 3m_{\ell}\pi s^{2}\sqrt{\lambda(s-4m^{2}_{\ell})}\{(|\mathcal{D}_{3}|^{2}+|\mathcal{D}_{9}|^{2})-(\mathcal{D}_{3}\mathcal{D}^{\ast}_{5}+\mathcal{D}_{5}\mathcal{D}^{\ast}_{3})
+(\mathcal{D}_{9}\mathcal{D}^{\ast}_{11}-\mathcal{D}_{11}\mathcal{D}^{\ast}_{9})\}\label{36}\\
\mathcal{H}^{T^{5}}_{6}&=&3m_{\ell}\pi s\sqrt{\lambda(s-4m^{2}_{\ell})}\{(M_{\Lambda_{b}}+M_{\Lambda})(\mathcal{D}_{3}\mathcal{D}^{\ast}_{1}+\mathcal{D}_{1}\mathcal{D}^{\ast}_{5}
-\mathcal{D}_{9}\mathcal{D}^{\ast}_{7})+(M_{\Lambda_{b}}-M_{\Lambda})(\mathcal{D}_{7}\mathcal{D}^{\ast}_{11}-\mathcal{D}_{11}\mathcal{D}^{\ast}_{7})\notag\\
&&-(M_{\Lambda_{b}}^{2}-M_{\Lambda}^{2})(|\mathcal{D}_{3}|^{2}+|\mathcal{D}_{9}|^{2})-(|\mathcal{D}_{1}|^{2}+|\mathcal{D}_{7}|^{2})\}\label{37}
 \end{eqnarray}
 where $\mathcal{D}_{1,3,5}=f_{1,2,3},\mathcal{D}_{2,4,6}=\frac{2m_{b}}{s}f^{T}_{1,2,3}$ and 
 $\mathcal{D}_{7,9,11}=g_{1,2,3},\mathcal{D}_{8,10,12}=\frac{2m_{b}}{s}g^{T}_{1,2,3}$
 \section{Numerical Analysis}
  In this section we will discuss the numerical analysis of the unpolarized and polarized CP violation asymmetries for $\Lambda_b\to \Lambda \ell^{+} \ell^{-}$ with $\ell = \mu, \tau$ decays.
  In order to see the imprints of the family non-universal $Z^{\prime}$ gauge boson on the said physical observables, first we have to summarize the numerical values of 
  various input parameters used in calculations such as masses of particles, life time, quark coupling CKM matrix etc., in Table I, while the values of Wilson coefficents
  are presented in Table-II. The most important input parameters which are important in any hadronic decays are the non perturbative quantities, called form factors, 
  and for the said decay we rely on light cone QCD sum rules approach \cite{Aliev2010}. The parametrization of the form factors $f_{1,2,3}$ , $g_{1,2,3}$, $f^{T}_{2,3}$ and $g^{T}_{2,3}$ are given by
\begin{equation}
f_i(q^{2})[g_i(q^{2})]=\frac{a}{1-q^{2}/m_{fit}^{2}}+\frac{b}{(1-q^{2}/m_{fit}^{2})^2},
\label{form-factor-sq}
\end{equation}
while the form factors $f^{T}_{1}$ and $g^{T}_{1}$ are of the form 
\begin{equation}
f^{T}_1[g^{T}_1]=\frac{c}{1-q^{2}/m_{fit}^{'2}}+\frac{c}{(1-q^{2}/m_{fit}^{''2})^{2}}.
\end{equation}
The numerical values of the light cone QCD sum rules form factors along with
the different fitting parameters \cite{Aliev2010} are summarized in Table III and IV.
\begin{table}[ht]
\caption{Default values of input parameters used in the calculations.}
\label{input}\centering
\begin{tabular}{c}
\hline\hline
$M_{\Lambda_b}=5.620$ GeV, $m_{b}=4.28$ GeV, $m_{s}=0.13$ GeV, \\
$m_{\mu}=0.105$ GeV, $m_{\tau}=1.77$ GeV, \\
$|V_{tb}V_{ts}^{\ast}|=45\times 10^{-3}$, $\alpha^{-1}=137$, $%
G_{F}=1.17\times 10^{-5}$ GeV$^{-2}$, \\
$\tau_{\Lambda_b}=1.383\times 10^{-12}$ sec, $M_{\Lambda}=1.115$ GeV. \\ \hline\hline
\end{tabular}%
\end{table}

\begin{table*}[ht]
\centering \caption{The Wilson coefficients $C_{i}^{\mu}$ at the
scale $\mu\sim m_{b}$ in the SM \cite{AAli}.}
\begin{tabular}{cccccccccc}
\hline\hline
$C_{1}$&$C_{2}$&$C_{3}$&$C_{4}$&$C_{5}$&$C_{6}$&$C_{7}$&$C_{9}$&$C_{10}$
\\ \hline
 \ \  1.107 \ \  & \ \  -0.248 \ \  & \ \  -0.011 \ \  & \ \  -0.026 \ \  & \ \  -0.007 \ \  & \ \  -0.031 \ \  & \ \  -0.313 \ \  & \ \  4.344 \ \  & \ \  -4.669 \ \  \\
\hline\hline
\end{tabular}
\label{wc table}
\end{table*}

\begin{table}
\caption{Fit parameters for $\Lambda_b \rightarrow \Lambda \ell^- \ell^- $ transition form
factors in full theory. Here only central value is given \cite{Aliev2010}}
\label{di-fit lambdab to lambda)}%
\begin{tabular}{ccccccc}
\hline\hline
&  & $\hspace{1 cm} a$ &  $\hspace{1 cm} b$ & $\hspace{1 cm} m_{fit}^{2}$ \\ \hline
& $f_1$ & $\hspace{1 cm}-0.046$ & $\hspace{1 cm} 0.368$ & $\hspace{1 cm} 39.10$ \\ \hline
& $f_2$ & $\hspace{1 cm}0.0046$ & $\hspace{1 cm} -0.017$ & $\hspace{1 cm} 26.37$ \\ \hline
& $f_3$ & $\hspace{1 cm}0.006$ & $\hspace{1 cm} -0.021$ & $\hspace{1 cm} 22.99$ \\ \hline
& $g_1$ & $\hspace{1 cm}-0.220$ & $\hspace{1 cm} 0.538$ & $\hspace{1 cm} 48.70$ \\ \hline
& $g_2$ & $\hspace{1 cm}0.005$ & $\hspace{1 cm} -0.018$ & $\hspace{1 cm} 26.93$ \\ \hline
& $g_3$ & $\hspace{1 cm}0.035$ & $\hspace{1 cm} -0.050$ & $\hspace{1 cm} 24.26$ \\ \hline
& $f^{T}_2$ & $\hspace{1 cm}-0.131$ & $\hspace{1 cm}0.426$ & $\hspace{1 cm} 45.70$ \\ \hline
& $f^{T}_3$ & $\hspace{1 cm}-0.046$ & $\hspace{1 cm} 0.102$ & $\hspace{1 cm} 28.31$ \\ \hline
& $g^{T}_2$ & $\hspace{1 cm}-0.369$ & $\hspace{1 cm} 0.664$ & $\hspace{1 cm} 59.37$ \\ \hline
& $f^{T}_2$ & $\hspace{1 cm}-0.026$ & $\hspace{1 cm} -0.075$ & $\hspace{1 cm} 23.73$ \\ 
\hline\hline
&  &  &  &  &  &
\end{tabular}%
\end{table}

\begin{table}
\caption{Fit parameters for $\Lambda_b \rightarrow \Lambda \ell^- \ell^- $ transition form
factors in full theory for $f^{T}_1$ and $g^{T}_1$ Here only central value is given \cite{Aliev2010}}
\label{di-fit lambdab to lambda1)}%
\begin{tabular}{ccccccc}
\hline\hline
&  & $\hspace{1 cm} c$ &  $\hspace{1 cm} m_{fit}^{'2}$ & $\hspace{1 cm} m_{fit}^{''2}$ \\ \hline
& $f^{T}_1$ & $\hspace{1 cm}-1.191$ & $\hspace{1 cm} 23.81$ & $\hspace{1 cm} 59.96$ \\ \hline
& $g^{T}_1$ & $\hspace{1 cm}-0.653$ & $\hspace{1 cm} 24.15$ & $\hspace{1 cm} 48.52$ \\ 
\hline\hline
&  &  &  &  &  &
\end{tabular}%
\end{table} 

Regarding to the couplings of family non universal $Z^{\prime}$-model there are some strong constraint from both inclusive and exclusive $B$ meson decays \cite{ConstrainedZPC1}.
The numerical values of quarks and leptons coupligns parameters of $Z^{\prime}$ model are given in Table V, where $\mathcal{S}1$ and $\mathcal{S}2$ represents the two different
fitting values for $B_s-\bar{B_{s}}$ mixing data by the
UTfit collaboration \cite{UTfit} and the numerical values of $\mathcal{S}3$ are chosen from \cite{cpv5, newcon4} and are also summarized in Table V.

\begin{table*}[ht]
\centering \caption{The numerical values of the $Z^\prime$
parameters \cite{ConstrainedZPC1,UTfit,cpv5,newcon4}.}
\begin{tabular}{cccccc}
\hline\hline \ \ &  $|\mathcal{B}_{sb}|\times10^{-3}$ \ \  & \ \ $
\phi_{sb}(in Degree)$ \ \  & \ \ $S_{LL}\times10^{-2}$ \ \  & \ \
$D_{LL}\times10^{-2}$
\\ \hline
 \ \  $\mathcal{S}1$ \ \  & \ \  $1.09\pm0.22$ \ \  & \ \  $-72\pm7$ \ \  & \ \  $-2.8\pm3.9$ \ \  & \ \  $-6.7\pm2.6$ \\
\ \  $\mathcal{S}2$ \ \  & \ \  $2.20\pm0.15$ \ \  & \ \  $-82\pm4$ \ \  & \ \  $-1.2\pm1.4$ \ \  & \ \  $-2.5\pm0.9$  \\
\ \  $\mathcal{S}3$ \ \  & \ \  $4.0\pm1.5$ \ \  & \ \  $150\pm10$ or $(-150\pm10)$ \ \  & \ \  $0.8$ \ \  & \ \  $-2.6$  \\
\hline\hline
\end{tabular}
\label{ZP table}
\end{table*}

It has been already mentioned that $\mathcal{B}_{sb}=|\mathcal{B}_{sb}|e^{-i\phi_{sb}}$ is the off diagonal left handed coupling of
$Z^\prime$ boson with quarks and $\phi_{sb}$ corresponds to a new weak phase, whereas
$S_{LL}$ and $D_{LL}$ represent the combination of left and
right handed couplings of $Z^\prime$ with the leptons [c.f. Eq. (\ref{C9C10})].
In order to fully scan the three scenarios, let us make a remark that with $D_{LL}\neq0$
depict the situation when the new physics comes only from the
modification in the Wilson coefficient $C_{10}$, while, the opposite
case, $S_{LL}\neq0$ , indicates that the new physics
is due to the change in
the Wilson coefficient $C_{9}$ [see Eq. (\ref{C9C10})].  In Figs. \ref{UnCPmu} to \ref{TrCPtau} the average CP violating asymmetries, after integration on $s$,
as a function of $S_{LL}$ and $D_{LL}$ are depicted. 
The different color codes along with the values of $Z^{\prime}$ parameters in scenario $\mathcal{S}1$ and $\mathcal{S}2$ are summarize in
Table VI. However for scenario $\mathcal{S}3$ the values of $Z^{\prime}$ parameters with different color codes are given in Eq. (\ref{Newcons}).
\begin{equation}
|\mathcal{B}_{sb}|=3\times 10^{-3}:\left\{
\begin{array}{c}

\phi _{sb}=-140^{\circ }\text{, Magenta Dot} \\
\phi _{sb}=-160^{\circ }\text{, Pink Dot}%
\end{array}%
\right. \text{; }|\mathcal{B}_{sb}|=5\times 10^{-3}:\left\{
\begin{array}{c}
\phi _{sb}=-140^{\circ }\text{, Green Dot} \\
\phi _{sb}=-160^{\circ }\text{, Red Dot}%
\end{array}%
\right. \label{Newcons}
\end{equation}

\begin{table*}[ht] \centering \caption{Color bands for the Figs. 1 $-$ 8 $\langle\mathcal{A_{CP}}\rangle$ and
$\langle\mathcal{A}^{i}_{CP}\rangle$ vs $S_{LL}$ and $D_{LL}$ for scenarios $\mathcal{S}1$ and $\mathcal{S}2$.}
\begin{tabular}{|c|c|c|c|c|} \cline{1-5}
\multicolumn{1}{|c|}{\multirow{2}{*}{Color Region}} &
\multicolumn{1}{|c|}{\multirow{2}{*}{$\phi_{sb}$}} &
\multicolumn{1}{|c|}{\multirow{2}{*}{$|\mathcal{B}_{sb}|\times 10^{-3}$}} &
\multicolumn{1}{|c|}{\multirow{1}{*}{$\langle\mathcal{A_{CP}}\rangle$ and
$\langle\mathcal{A}^{i}_{CP}\rangle$ vs $S_{LL}$}} &
\multicolumn{1}{|c|}{\multirow{1}{*}{$\langle\mathcal{A_{CP}}\rangle$ and
$\langle\mathcal{A}^{i}_{CP}\rangle$ vs $D_{LL}$}} \\
\multicolumn{1}{|c|}{\multirow{1}{*}{}} &
\multicolumn{1}{|c|}{\multirow{1}{*}{}} &
\multicolumn{1}{|c|}{\multirow{1}{*}{}} &
\multicolumn{1}{|c|}{\multirow{1}{*}{$D_{LL}\times 10^{-2}$}} &
\multicolumn{1}{|c|}{\multirow{1}{*}{$S_{LL}\times 10^{-2}$}} \\
\cline{1-5}\multicolumn{1}{|c|}{\multirow{2}{*}{Blue}} &
\multicolumn{1}{|c|}{\multirow{1}{*}{$-79^\circ$}} &
\multicolumn{1}{|c|}{\multirow{2}{*}{+1.31}} &
\multicolumn{1}{|c|}{\multirow{2}{*}{-4.1}} &
\multicolumn{1}{|c|}{\multirow{2}{*}{+1.1}} \\
\multicolumn{1}{|c|}{\multirow{1}{*}{}} &
\multicolumn{1}{|c|}{\multirow{1}{*}{$-65^\circ$}} &
\multicolumn{1}{|c|}{\multirow{1}{*}{}} &
\multicolumn{1}{|c|}{\multirow{1}{*}{}} &
\multicolumn{1}{|c|}{\multirow{1}{*}{}} \\
\cline{1-5} \multicolumn{1}{|c|}{\multirow{2}{*}{Red}} &
\multicolumn{1}{|c|}{\multirow{1}{*}{$-79^\circ$}} &
\multicolumn{1}{|c|}{\multirow{2}{*}{+1.31}} &
\multicolumn{1}{|c|}{\multirow{2}{*}{-9.3}} &
\multicolumn{1}{|c|}{\multirow{2}{*}{-6.7}} \\
\multicolumn{1}{|c|}{\multirow{1}{*}{}} &
\multicolumn{1}{|c|}{\multirow{1}{*}{$-65^\circ$}} &
\multicolumn{1}{|c|}{\multirow{1}{*}{}} &
\multicolumn{1}{|c|}{\multirow{1}{*}{}} &
\multicolumn{1}{|c|}{\multirow{1}{*}{}} \\
\cline{1-5} \multicolumn{1}{|c|}{\multirow{2}{*}{Yellow}} &
\multicolumn{1}{|c|}{\multirow{1}{*}{$-79^\circ$}} &
\multicolumn{1}{|c|}{\multirow{2}{*}{+0.87}} &
\multicolumn{1}{|c|}{\multirow{2}{*}{-4.1}} &
\multicolumn{1}{|c|}{\multirow{2}{*}{+1.1}} \\
\multicolumn{1}{|c|}{\multirow{1}{*}{}} &
\multicolumn{1}{|c|}{\multirow{1}{*}{$-65^\circ$}} &
\multicolumn{1}{|c|}{\multirow{1}{*}{}} &
\multicolumn{1}{|c|}{\multirow{1}{*}{}} &
\multicolumn{1}{|c|}{\multirow{1}{*}{}} \\
\cline{1-5} \multicolumn{1}{|c|}{\multirow{2}{*}{Gray}} &
\multicolumn{1}{|c|}{\multirow{1}{*}{$-79^\circ$}} &
\multicolumn{1}{|c|}{\multirow{2}{*}{+0.87}} &
\multicolumn{1}{|c|}{\multirow{2}{*}{-9.3}} &
\multicolumn{1}{|c|}{\multirow{2}{*}{-6.7}} \\
\multicolumn{1}{|c|}{\multirow{1}{*}{}} &
\multicolumn{1}{|c|}{\multirow{1}{*}{$-65^\circ$}} &
\multicolumn{1}{|c|}{\multirow{1}{*}{}} &
\multicolumn{1}{|c|}{\multirow{1}{*}{}} &
\multicolumn{1}{|c|}{\multirow{1}{*}{}} \\
\cline{1-5} \multicolumn{1}{|c|}{\multirow{2}{*}{Green}} &
\multicolumn{1}{|c|}{\multirow{1}{*}{$-86^\circ$}} &
\multicolumn{1}{|c|}{\multirow{2}{*}{+2.35}} &
\multicolumn{1}{|c|}{\multirow{2}{*}{-1.6}} &
\multicolumn{1}{|c|}{\multirow{2}{*}{+0.2}} \\
\multicolumn{1}{|c|}{\multirow{1}{*}{}} &
\multicolumn{1}{|c|}{\multirow{1}{*}{$-78^\circ$}} &
\multicolumn{1}{|c|}{\multirow{1}{*}{}} &
\multicolumn{1}{|c|}{\multirow{1}{*}{}} &
\multicolumn{1}{|c|}{\multirow{1}{*}{}} \\
\cline{1-5} \multicolumn{1}{|c|}{\multirow{2}{*}{Orange}} &
\multicolumn{1}{|c|}{\multirow{1}{*}{$-86^\circ$}} &
\multicolumn{1}{|c|}{\multirow{2}{*}{+2.35}} &
\multicolumn{1}{|c|}{\multirow{2}{*}{-3.4}} &
\multicolumn{1}{|c|}{\multirow{2}{*}{-2.6}} \\
\multicolumn{1}{|c|}{\multirow{1}{*}{}} &
\multicolumn{1}{|c|}{\multirow{1}{*}{$-78^\circ$}} &
\multicolumn{1}{|c|}{\multirow{1}{*}{}} &
\multicolumn{1}{|c|}{\multirow{1}{*}{}} &
\multicolumn{1}{|c|}{\multirow{1}{*}{}} \\
\cline{1-5} \multicolumn{1}{|c|}{\multirow{2}{*}{Pink}} &
\multicolumn{1}{|c|}{\multirow{1}{*}{$-86^\circ$}} &
\multicolumn{1}{|c|}{\multirow{2}{*}{+2.05}} &
\multicolumn{1}{|c|}{\multirow{2}{*}{-1.6}} &
\multicolumn{1}{|c|}{\multirow{2}{*}{+0.2}} \\
\multicolumn{1}{|c|}{\multirow{1}{*}{}} &
\multicolumn{1}{|c|}{\multirow{1}{*}{$-78^\circ$}} &
\multicolumn{1}{|c|}{\multirow{1}{*}{}} &
\multicolumn{1}{|c|}{\multirow{1}{*}{}} &
\multicolumn{1}{|c|}{\multirow{1}{*}{}} \\
\cline{1-5} \multicolumn{1}{|c|}{\multirow{2}{*}{Purple}} &
\multicolumn{1}{|c|}{\multirow{1}{*}{$-86^\circ$}} &
\multicolumn{1}{|c|}{\multirow{2}{*}{+2.05}} &
\multicolumn{1}{|c|}{\multirow{2}{*}{-3.4}} &
\multicolumn{1}{|c|}{\multirow{2}{*}{-2.6}} \\
\multicolumn{1}{|c|}{\multirow{1}{*}{}} &
\multicolumn{1}{|c|}{\multirow{1}{*}{$-78^\circ$}} &
\multicolumn{1}{|c|}{\multirow{1}{*}{}} &
\multicolumn{1}{|c|}{\multirow{1}{*}{}} &
\multicolumn{1}{|c|}{\multirow{1}{*}{}} \\
\cline{1-5}
\end{tabular}
\end{table*}

\textbf{Unpolarized $CP$ violation asymmetry:}
\begin{itemize}
\item Figs.\ref{UnCPmu} and \ref{UnCPmutau} represents the unpolarized $CP$ violation asymmetries for the decay $\Lambda_b\to\Lambda\mu^+\mu^-(\tau^+\tau^-)$ as a function
of $D_{LL}$ and $S_{LL}$ respectively. In standard model $CP$ violation asymmetry is zero hence the non-zero value will give us a clue of 
physics beyond the standard model which is commonly known as New Physics (NP). It is evident from Eq.(\ref{17}) that the $\mathcal{A}_{CP}$
is proportional to $Z^{\prime}$ parameters which comes through the imaginary part of the Wilson coefficents as well as of weak
phase $\phi_{sb}$ which conceals in $\Lambda_{sb}$(c.f.Eq.(\ref{lamb})). Therefore the dependence on new weak phase $\phi_{sb}$
is expected and is evident from Figs.\ref{UnCPmu} and \ref{UnCPmutau} where band in each case depicits the variation of new phase $\phi_{sb}$
in respective scenarios. In Fig.(\ref{UnCPmu}) $\mathcal{A}_{CP}$ is plotted vs $D_{LL}$ by changing the values of $S_{LL}$, $\phi_{sb}$
and $B_{sb}$. In case of $\mu$'s as a final state leptons, the value of $\mathcal{A}_{CP}$ is positive in both scenarios $\mathcal{S}1$
and $\mathcal{S}2$ but for positive values of $S_{LL}$. However, the value of $\mathcal{A}_{CP}$ reaches to $-0.05$ when
$D_{LL}=-1.6\times10^{-2}$ and corresponding $S_{LL}=-6.7\times10^{-2}$ depicited by red band. Similarly for the case of $\tau$'s as final state leptons,
the value of $\mathcal{A}_{CP}$ is positive in both scenarios for positive values of $S_{LL}$. However, the value of
$\mathcal{A}_{CP}$ is around $-0.08$ for $D_{LL}=-1.6\times10^{-2}$ and $S_{LL}=-6.7\times10^{-2}$ shown by red bands.
\begin{figure}[tbp]
\begin{tabular}{cc}
\includegraphics[scale=0.55]{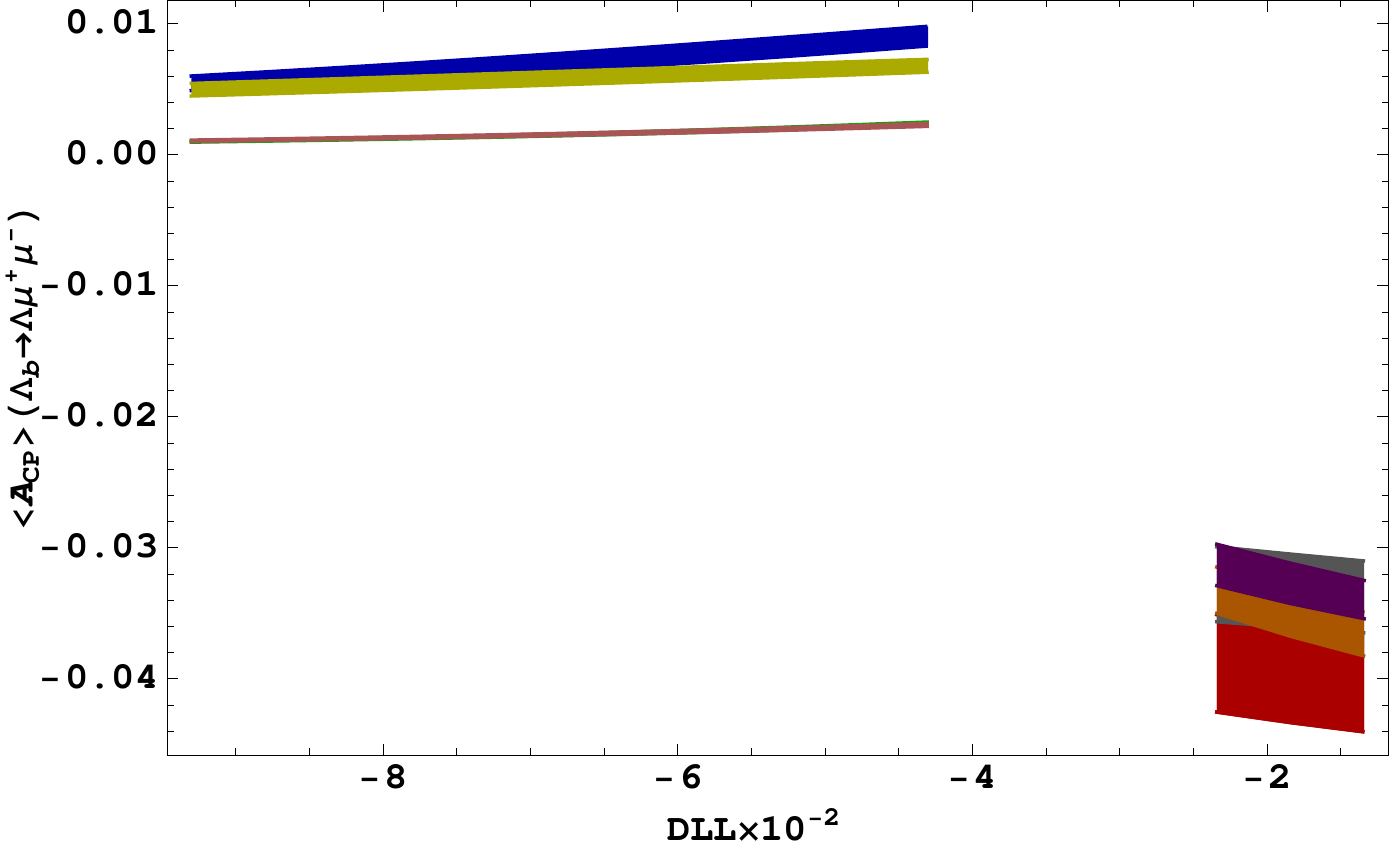}  \put (-100,150){(a)} & %
\includegraphics[scale=0.55]{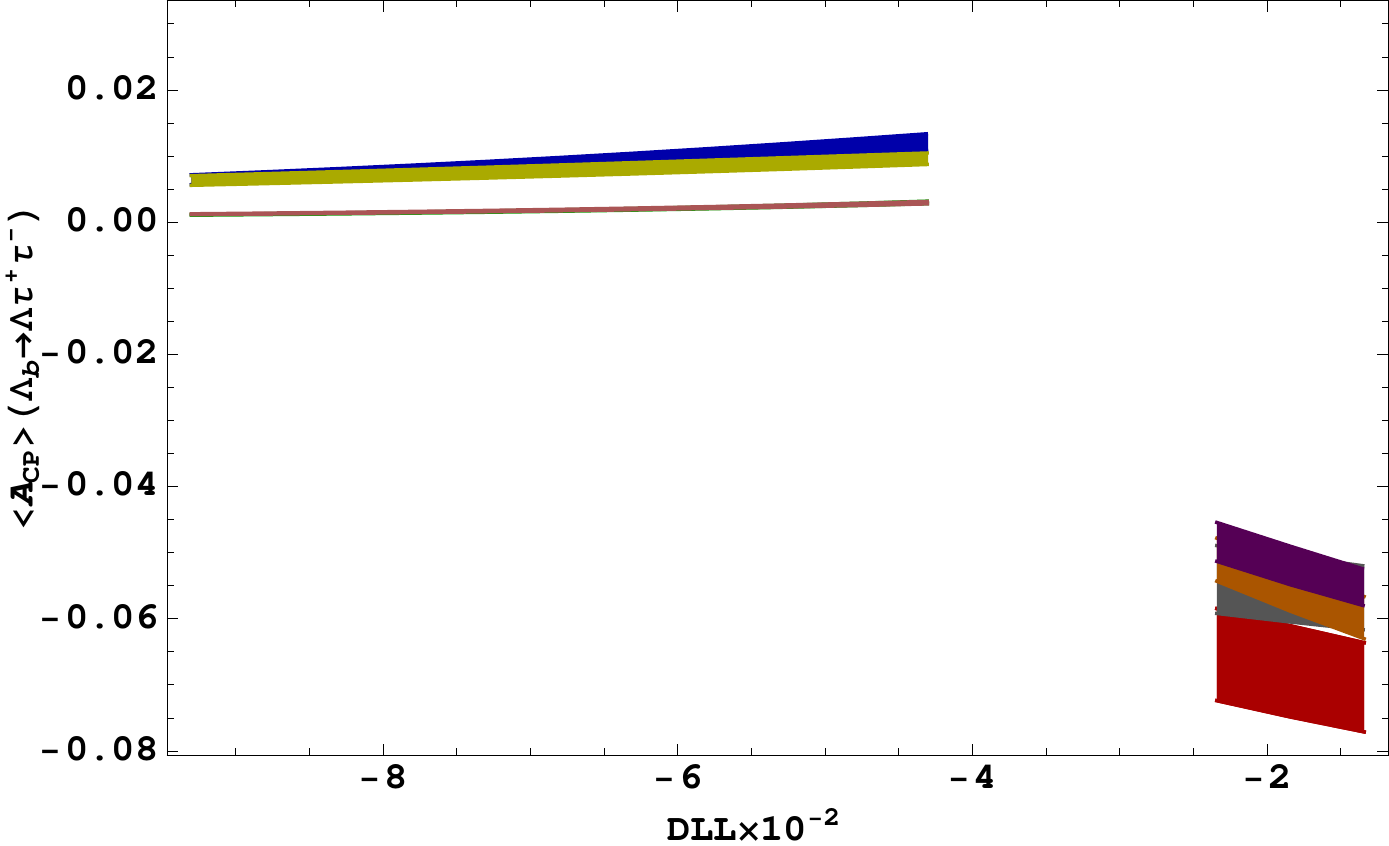}  \put (-100,150){(b)}%
\end{tabular}%
\caption{Unpolarized $CP$ violation asymmetry $\mathcal{A}_{CP}$ as
function of $D_{LL}$ for $\Lambda_{b}\to \Lambda \mu^+ \mu^- (\tau^+\tau^-)$ for scenarios $\mathcal{S}1$,
$\mathcal{S}2$ and $\mathcal{S}3$ . The red, blue, grey and yellow bands correspnds to $\mathcal{S}1$. Green, orange, pink and purple band correspnds to $\mathcal{S}2$. 
 The dots of different colors corresponds to $\mathcal{S}3$. The band in each case depicts the variations of $\phi_{sb}$ in respective
scenario.} \label{UnCPmu}
\end{figure}
\begin{figure}[tbp]
\begin{tabular}{cc}
\includegraphics[scale=0.55]{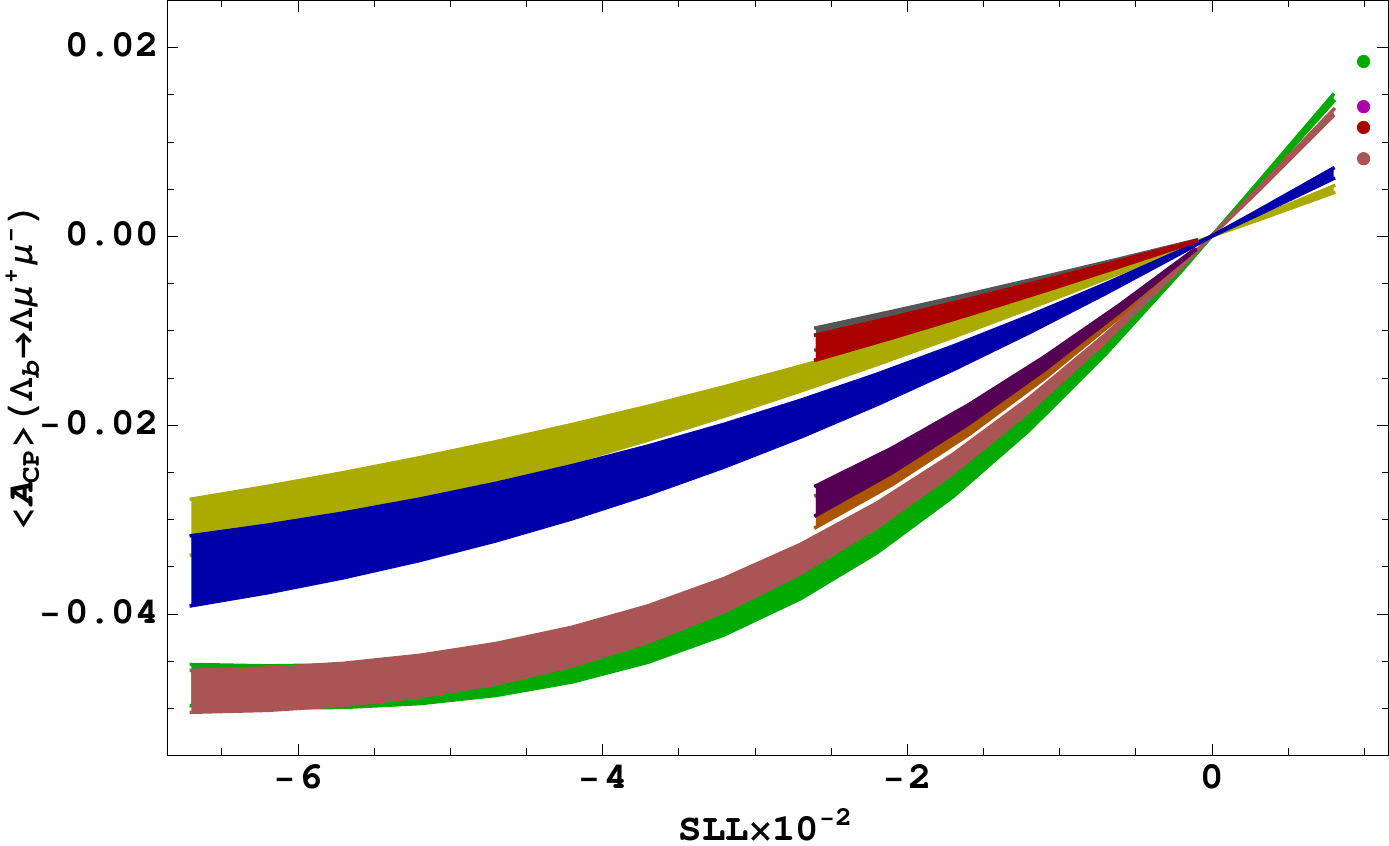}  \put (-100,150){(a)} & %
\includegraphics[scale=0.55]{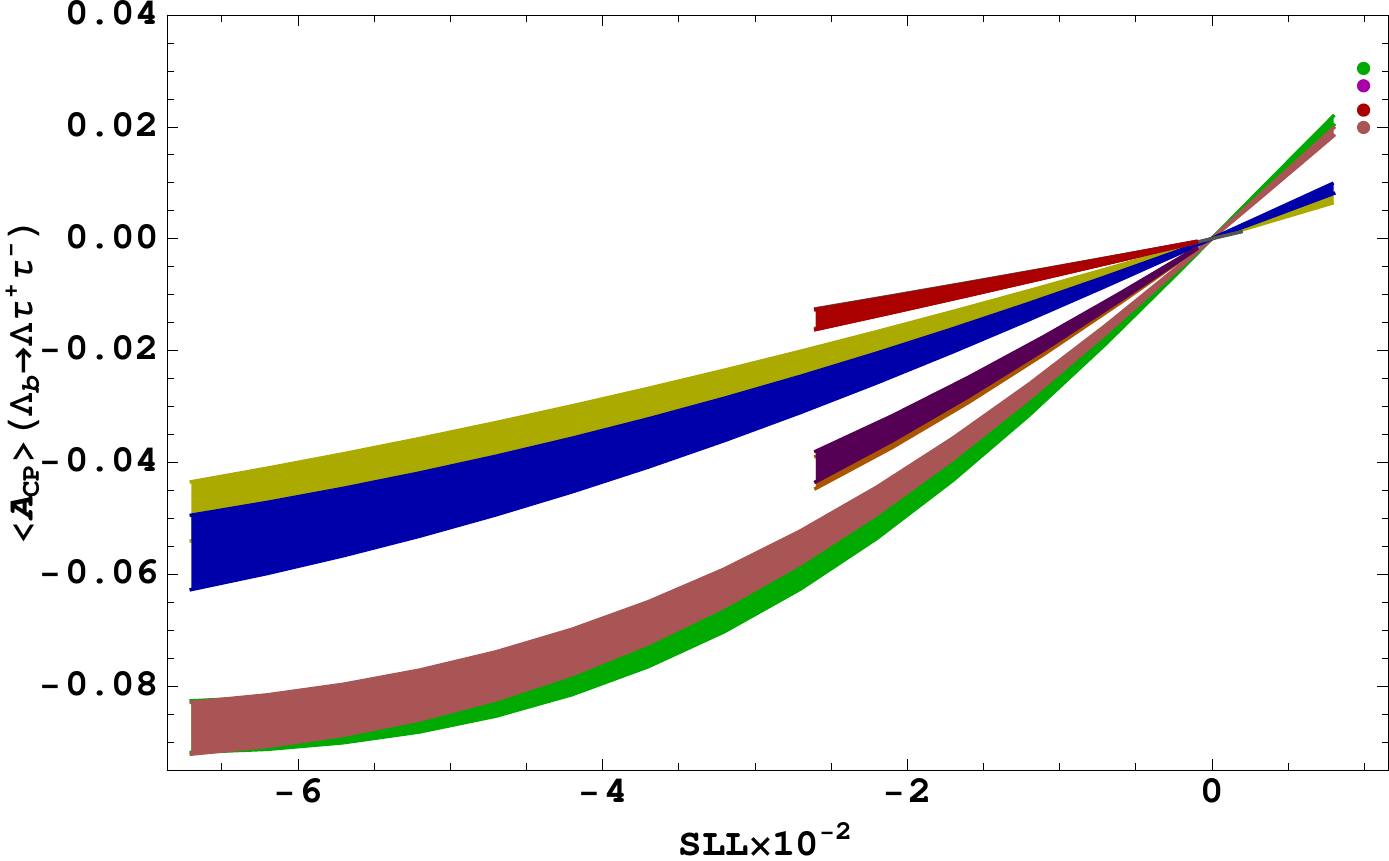}  \put (-100,150){(b)}%
\end{tabular}%
\caption{Unpolarized $CP$ violation asymmetry $\mathcal{A}_{CP}$ as
function of $S_{LL}$ for $\Lambda_{b}\to \Lambda \mu^+ \mu^- (\tau^+\tau^-)$ for scenarios $\mathcal{S}1$, $\mathcal{S}2$ and
$\mathcal{S}3$. The color and band description is same as in Fig. \ref{UnCPmu}.} \label{UnCPmutau}
\end{figure}
\item Fig. \ref{UnCPmutau} presents the behavior of $\mathcal{A}_{CP}$ with $S_{LL}$ by varying the values of $D_{LL},
\phi_{sb}$ and $\mathcal{B}_{sb}$ in the range given in Table V. It can be immediately noticed that in case of $\mu$'s the value is small compared
to the case when $\tau$'s are final state leptons. In both cases $\mathcal{A}_{CP}$ is an increasing function of the $S_{LL}$.
In $\Lambda_b \to \Lambda \tau^+ \tau^-$ the value of unpolarized $CP$ asymmetry is around $-0.1$  when $S_{LL}=-6.7 \times 10^{-2}$.

The values of unpolarized $CP$ violation asymmetries in scenario $\mathcal{S}3$ for $\Lambda_b \to \Lambda \mu^{+}\mu^{-}$ and $\Lambda_b \to \Lambda  \tau^{+}\tau^{-}$ are shown by different color dots in Figs. \ref{UnCPmutau}(a) and \ref{UnCPmutau}(b), respectively. It can be noticed that the value of unpolarized $CP$ violation asymmetry is positive maximum in this scenario when $\phi_{sb}=-140^{\circ}$, $|\mathcal{B}_{sb}|=5\times 10^{-3}$ and it is depicted by the green dot in these figures. Irrespective of the negative or positive values of the new weak phase $(\phi_{sb})$ the value of $CP$ asymmetry remains positive for all values of $\mathcal{B}_{sb}$ which is an entirely distinctive feature compared to the first two scenarios where $CP$ asymmetry climb from negative to positive value. 
\end{itemize}
\textbf{Longitudinal polarized $CP$ violation asymmetry:}
\begin{itemize}
\item The longitudinal polarized $CP$ violation asymmetry $\mathcal{A}^L_{CP}$ is plotted
in Figs. \ref{LnCPmu} and \ref{LnCPtau}. From Eq. (\ref{23}) it can be noticed that $\mathcal{Q}^{L}$ is proportional to the
imaginary part of the combination of Wilson coefficients which involve $C_{7}$, $C_{9}$ and $C_{10}$ both in the SM as well as in the $Z^{\prime}$ model. Even though, the Wilson coefficient $C_7$ does not get contribution from the $Z^{\prime}$, but the change in the Wilson coefficients $C_{9}$ and $C_{10}$ due to the parameters of $Z^\prime$ model will make the 
 $\mathcal{A}^L_{CP}$ sensitive to the change arising due to extra neutral boson $Z^\prime$.  In Fig. \ref{LnCPmu}(a) and \ref{LnCPmu}(b), we have plotted the $\mathcal{A}^L_{CP}$ vs $D_{LL}$ by fixing the values of $S_{LL}$ and other $Z^{\prime}$ parameters in the range given in Table V. We can see that the value of $\mathcal{A}^L_{CP}$
increases from $0.008$ to $0.053$ when $\mu$'s are the final state leptons and from $0.004$ to $0.030$
in case of $\tau$'s as final state leptons which can be visualized from the colour bands that corresponds to scenario $\mathcal{S}1(\mathcal{S}2)$.
The situation when the longitudinal polarized $CP$ violation asymmetry
is plotted with $S_{LL}$ by taking other parameters in the range given in Table V and it is displayed in Fig.\ref{LnCPtau}. Here we can see that it is an increasing function of $S_{LL}$ where in $\mathcal{S}1$ and $\mathcal{S}2$ the value increase from $0.01(0.005)$ to $0.05(0.025)$ when we have $\mu's(\tau's)$ as final state leptons and it is clearly visible from the red(pink) band. It can also be seen in Fig. \ref{LnCPtau}a, that value of the longitudinal polarized $CP$ violation asymmetry in scenario $\mathcal{S}3$ is much suppressed when we have $\mu$'s as final state leptons. However, in case of the $\tau$'s the value of the longitudinal $CP$ violation asymmetry is around $0.030$ when $\phi_{sb}=-140^{\circ}$ and $|\mathcal{B}_{sb}|=5\times 10^{-3}$. It is shown with the green dot in Fig. \ref{LnCPtau}b.  It can be noticed that  the value of longitudinal polarized $CP$ violation asymmetry in
 $\Lambda_{b}\to \Lambda \tau^{+}\tau^{-}$ is significantly different from its value in the $\mathcal{S}1$ and $\mathcal{S}2$. Hence, by measuring
$\mathcal{A}^L_{CP}$ one can not only segregate the NP coming through the $Z^{\prime}$ boson but can also distinguish the three scenarios.
\end{itemize}\begin{figure}[tbp]
\begin{tabular}{cc}
\includegraphics[scale=0.55]{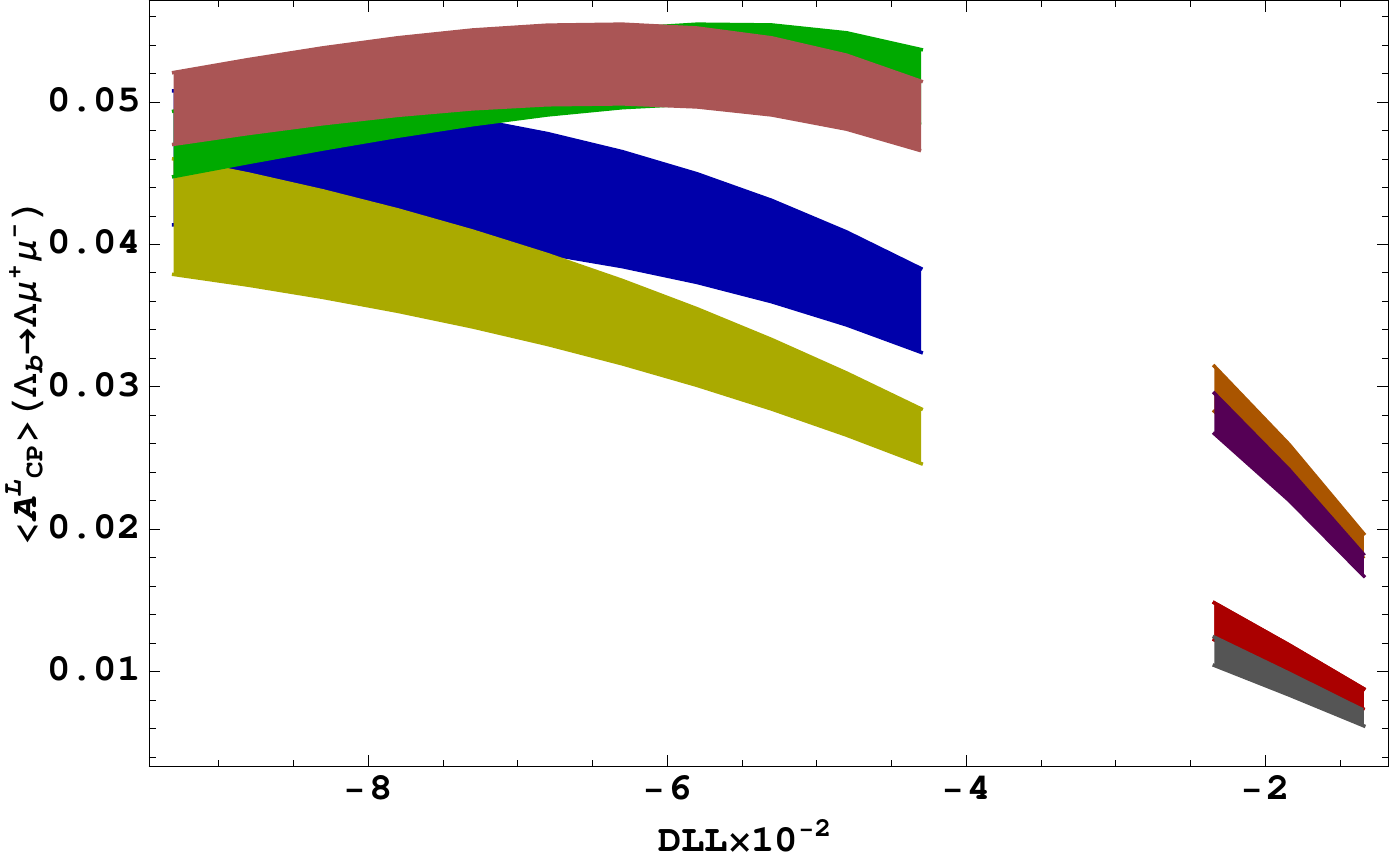}  \put (-100,150){(a)} & %
\includegraphics[scale=0.55]{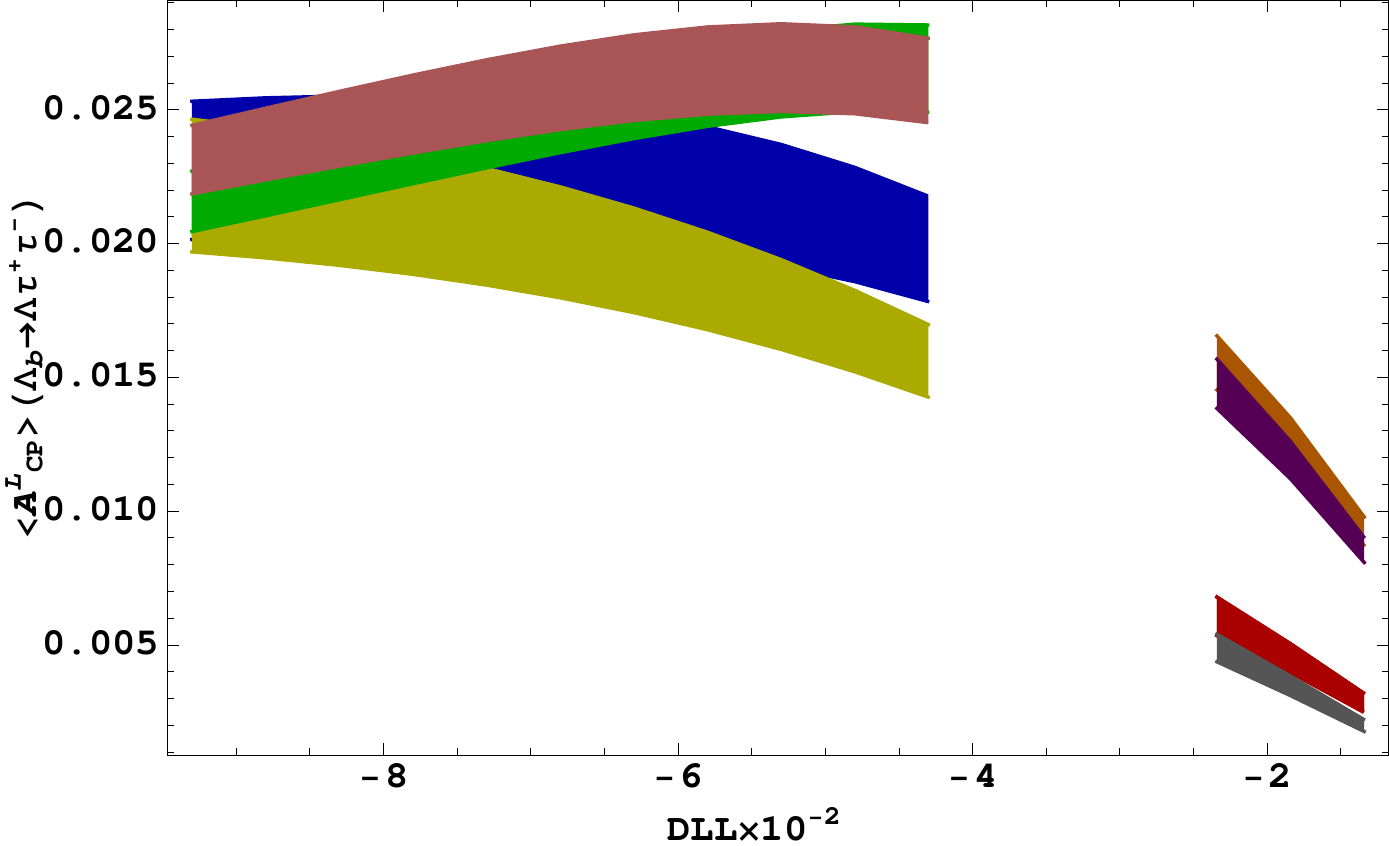}  \put (-100,150){(b)}%
\end{tabular}%
\caption{Longitudinally polarized $CP$ violation asymmetry $\mathcal{A}^{L}_{CP}$ as
function of $D_{LL}$ for $\Lambda_{b}\to \Lambda \mu^+ \mu^- (\tau^+\tau^-)$ for scenarios $\mathcal{S}1$,
$\mathcal{S}2$ and $\mathcal{S}3$ .The color and band description is same as in Fig. \ref{UnCPmu}.} \label{LnCPmu}
\end{figure}
\begin{figure}[tbp]
\begin{tabular}{cc}
\includegraphics[scale=0.55]{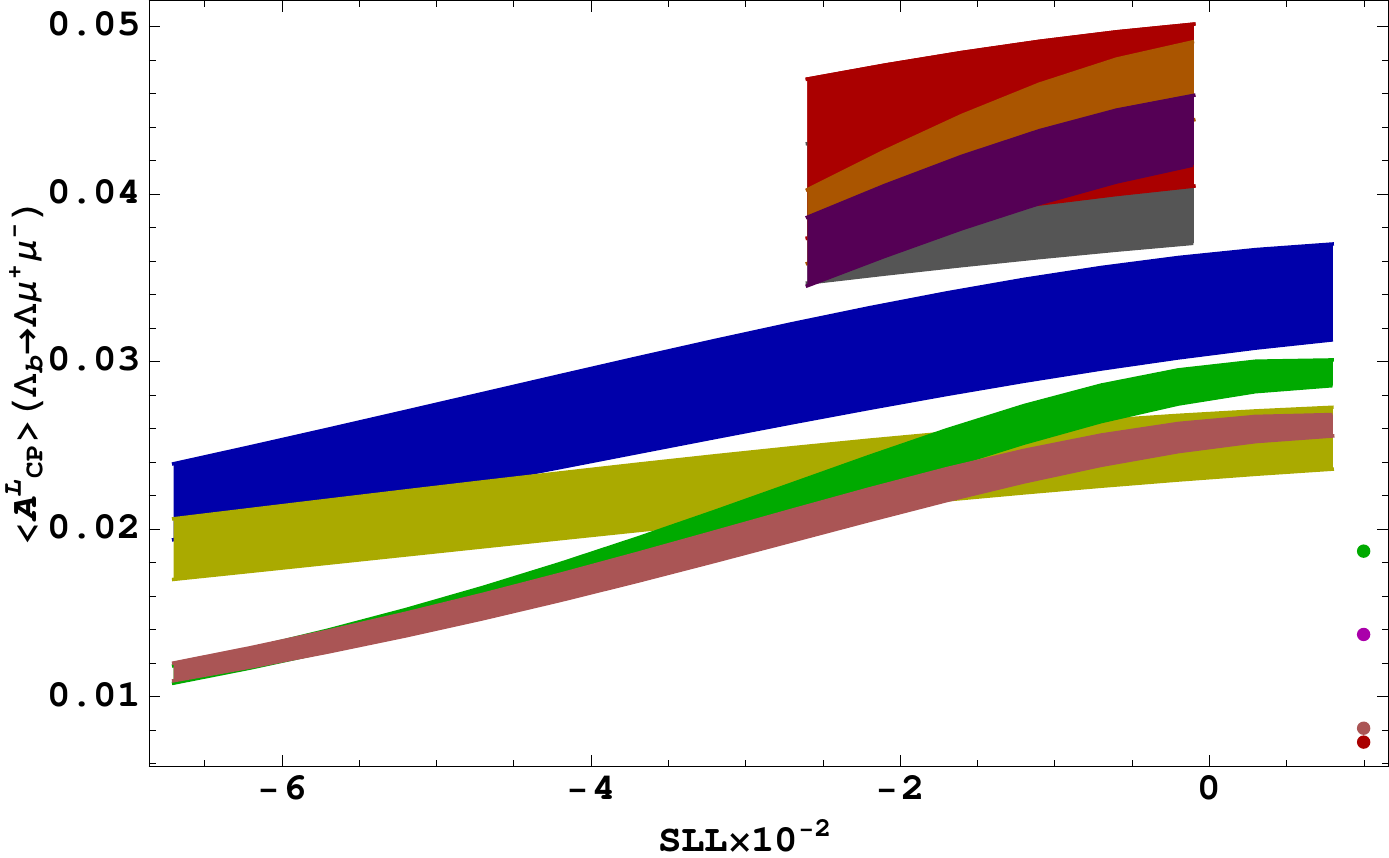}  \put (-100,150){(a)} & %
\includegraphics[scale=0.55]{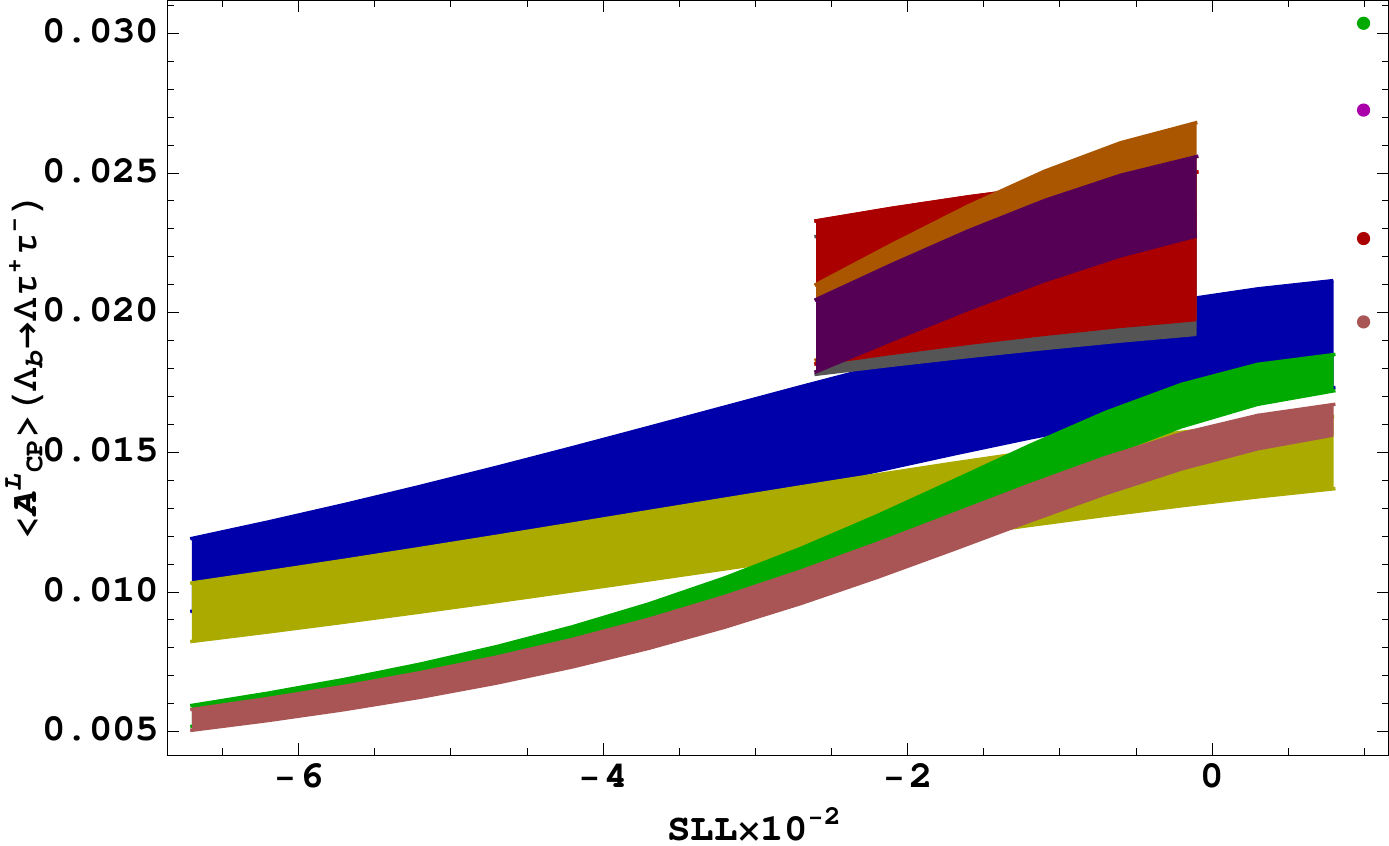}  \put (-100,150){(b)}%
\end{tabular}%
\caption{Longitudinal polarized $CP$ violation asymmetry $\mathcal{A}^{L}_{CP}$ as
function of $S_{LL}$ for $\Lambda_{b}\to \Lambda \mu^+ \mu^- (\tau^+\tau^-)$ for scenarios $\mathcal{S}1$,
$\mathcal{S}2$ and $\mathcal{S}3$. The color and band description is same as in Fig. \ref{UnCPmu}.} \label{LnCPtau}
\end{figure}

\textbf{Normal polarized $CP$ violation asymmetry:}
\begin{itemize}
\item Contrary to the  $\mathcal{A}_{CP}$ and $\mathcal{A}^{L}_{CP}$, the
normal polarized $CP$ violation asymmetry $(\mathcal{A}^{N}_{CP})$ is an order of magnitude smaller in case of $\mu$'s compared to the $\tau$'s as final state leptons.
By looking at the Eq. (\ref{26}). The $\mathcal{A}^{N}_{CP}$ comes from the function
$\mathcal{Q}^{N}$ which contains $\mathcal{H}_{1} ..., \mathcal{H}_6$. In Eqs. (\ref{27a} - \ref{30a}) it is clear
that these asymmetries are proportional to the lepton mass and their suppression in case of muon is obvious and Figs. \ref{NoCPmu}(a) and
\ref{NoCPtau}(a) depict this fact.
Coming to the Figs. \ref{NoCPmu}(b) and \ref{NoCPtau}(b) we can see that the
$\mathcal{A}^{N}_{CP}$ is very sensitive to the parameters of $Z^{\prime}$ both in
the $\mathcal{S}1$ and $\mathcal{S}2$. 
In Fig. \ref{NoCPmu}(b), the value of $\mathcal{A}^{N}_{CP}$ decreases from $0.040$ to $0.018$ in the parameter range of $Z^{\prime}$
in $\mathcal{S}1$ and from $0.048$ to $0.028$  in $\mathcal{S}2$.
The situation remains the same as in Fig. \ref{NoCPmu}  when $\mathcal{A}^{N}_{CP}$ is plotted with $S_{LL}$ in Figs. \ref{NoCPtau}(a) and \ref{NoCPtau}b. It can also be noted that average value of the $\mathcal{A}^{N}_{CP}$ increases from $0.020$ to $0.045$ in $\mathcal{S}1$ and $0.035$ to $0.05$ in  $\mathcal{S}2$. 

What comes out to be more interesting is the impact of parametric space of scenario $\mathcal{S}3$ in case of $\mu$'s and final state leptons. In this scenario, the value of the normal $CP$ violation asymmetry in $\Lambda_{b} \to \Lambda \mu^+ \mu^-$ is an order of magnitude larger than the corresponding values in $\mathcal{S}1$ and $\mathcal{S}2$.  Here, the maximum value is $0.018$ (the green dot) when $\phi_{sb}=-140^{\circ}$ and $|\mathcal{B}_{sb}|=5\times 10^{-3}$. While in case of $\tau$'s as final state leptons the order of asymmetries remains  the same
as in $\mathcal{S}1$ and $\mathcal{S}2$.
\end{itemize}

\textbf{Transverse polarized $CP$ violation asymmetry:}
\begin{itemize}
\item Just like the normal polarized $CP$ violation asymmetry, the
different terms in transverse polarized $CP$ violation asymmetry $\mathcal{A}^{T}_{CP}$ are also $m_{l}$ suppressed which is
visible from $\mathcal{H}$'s appearing in the function $\mathcal{Q}^T$ in Eq. (\ref{30b}). The graphs given in Figs. \ref{TrCPmu}(a) and \ref{TrCPtau}(a)
depict the fact that in the presence of NP, the maximum value of  $\mathcal{A}^{T}_{CP}$ is around $0.016$ (shown by the green band) in $\Lambda_{b} \rightarrow \Lambda \mu^+ \mu^-$,
 while Figs. \ref{TrCPmu}(b) and \ref{TrCPtau}(b)
are shown that in case of the $\tau$'s as final state
leptons the value of the $\mathcal{A}^{T}_{CP}$ reaches upto $0.08$ in certain parametric space of the $Z^{\prime}$ scenario $\mathcal{S}1$.

By varying the $Z^{\prime}$ parameters in the range given in Eq. (\ref{Newcons}) the trend of transverse $CP$ violation asymmetry is shown by different colors of dots in Fig.8. In case of $\mu$'s as final state leptons, we can see that for $\phi_{sb}=-140^{\circ}, |\mathcal{B}_{sb}|=5\times 10^{-3}$ in scenario $\mathcal{S}3$ the value of transverse polarized $CP$ violation asymmetry is slightly higher than the first two scenarios (shown by the green dot). However, in   $\Lambda_{b} \rightarrow \Lambda \tau^+ \tau^-$ decay the effects coming through the parametric space of $\mathcal{S}3$ are smaller than that of the first two scenarios. 
\end{itemize}

\begin{figure}[tbp]
\begin{tabular}{cc}
\includegraphics[scale=0.55]{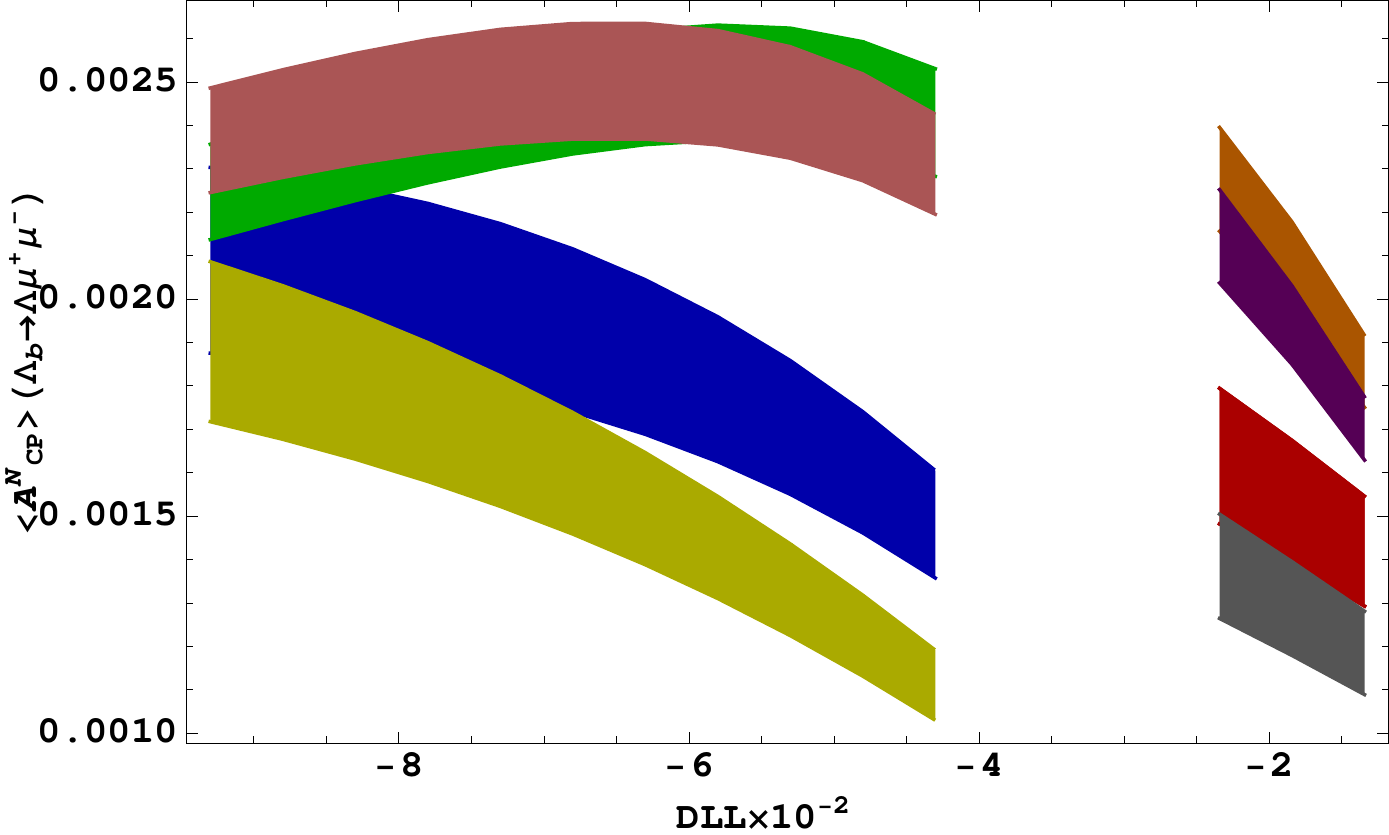}  \put (-100,150){(a)} & %
\includegraphics[scale=0.55]{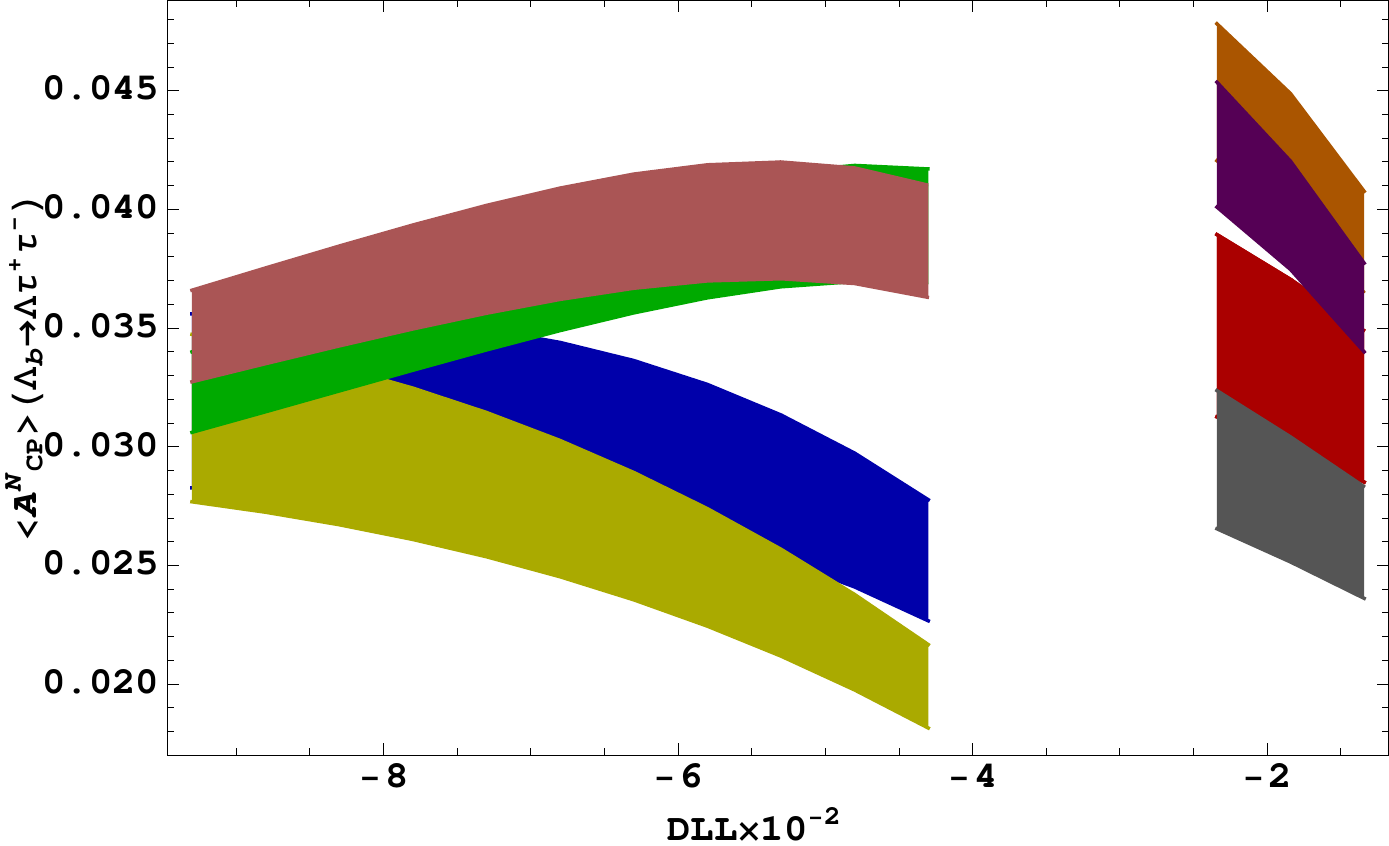}  \put (-100,150){(b)}%
\end{tabular}%
\caption{Normal polarized $CP$ violation asymmetry $\mathcal{A}^{N}_{CP}$ as
function of $D_{LL}$ for  $\Lambda_{b}\to \Lambda \mu^+ \mu^- (\tau^+\tau^-)$ for scenarios $\mathcal{S}1$,
$\mathcal{S}2$ and $\mathcal{S}3$. The color and band description is same as in Fig. \ref{UnCPmu}.} \label{NoCPmu}
\end{figure}
\begin{figure}[tbp]
\begin{tabular}{cc}
\includegraphics[scale=0.55]{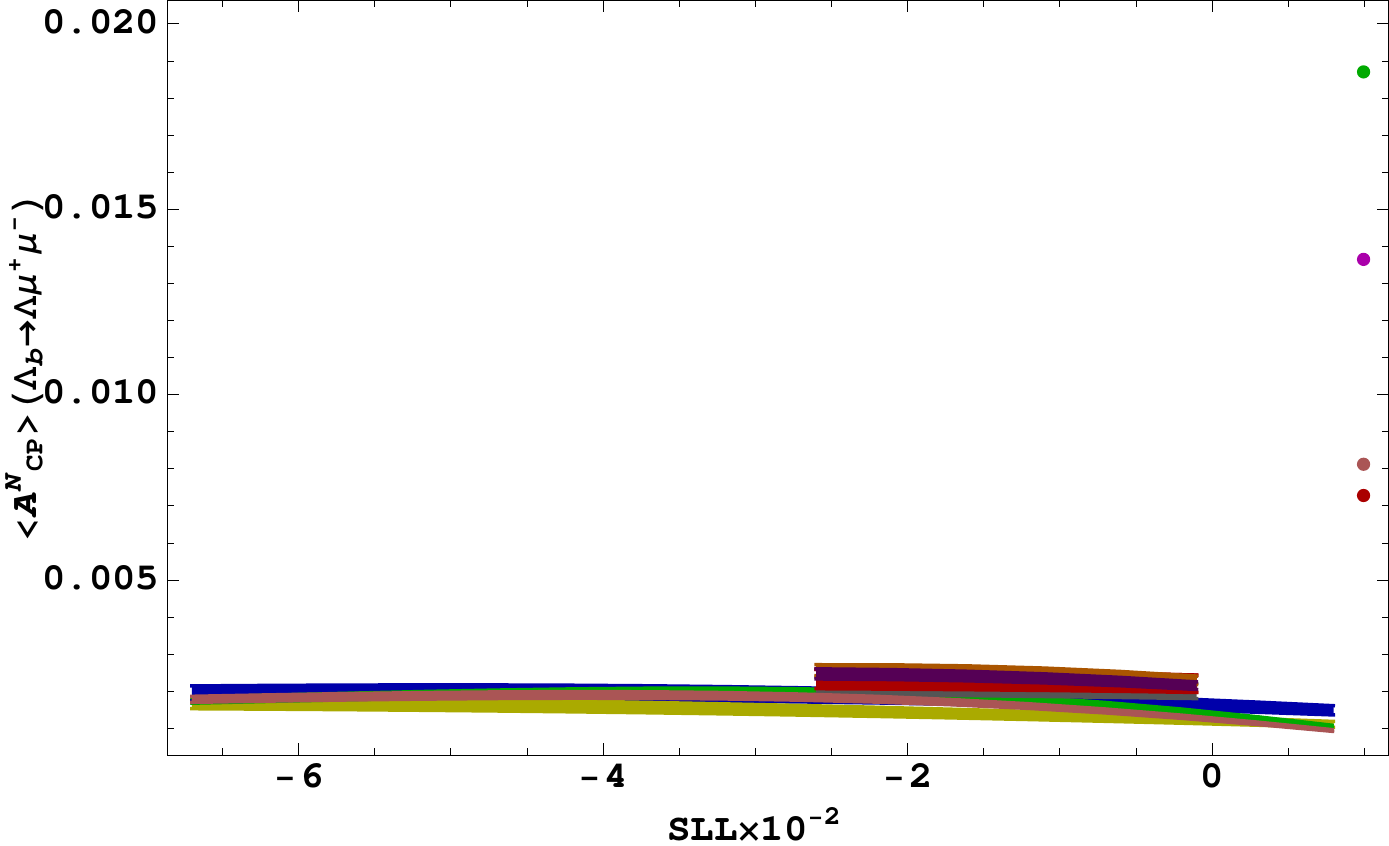}  \put (-100,150){(a)} & %
\includegraphics[scale=0.55]{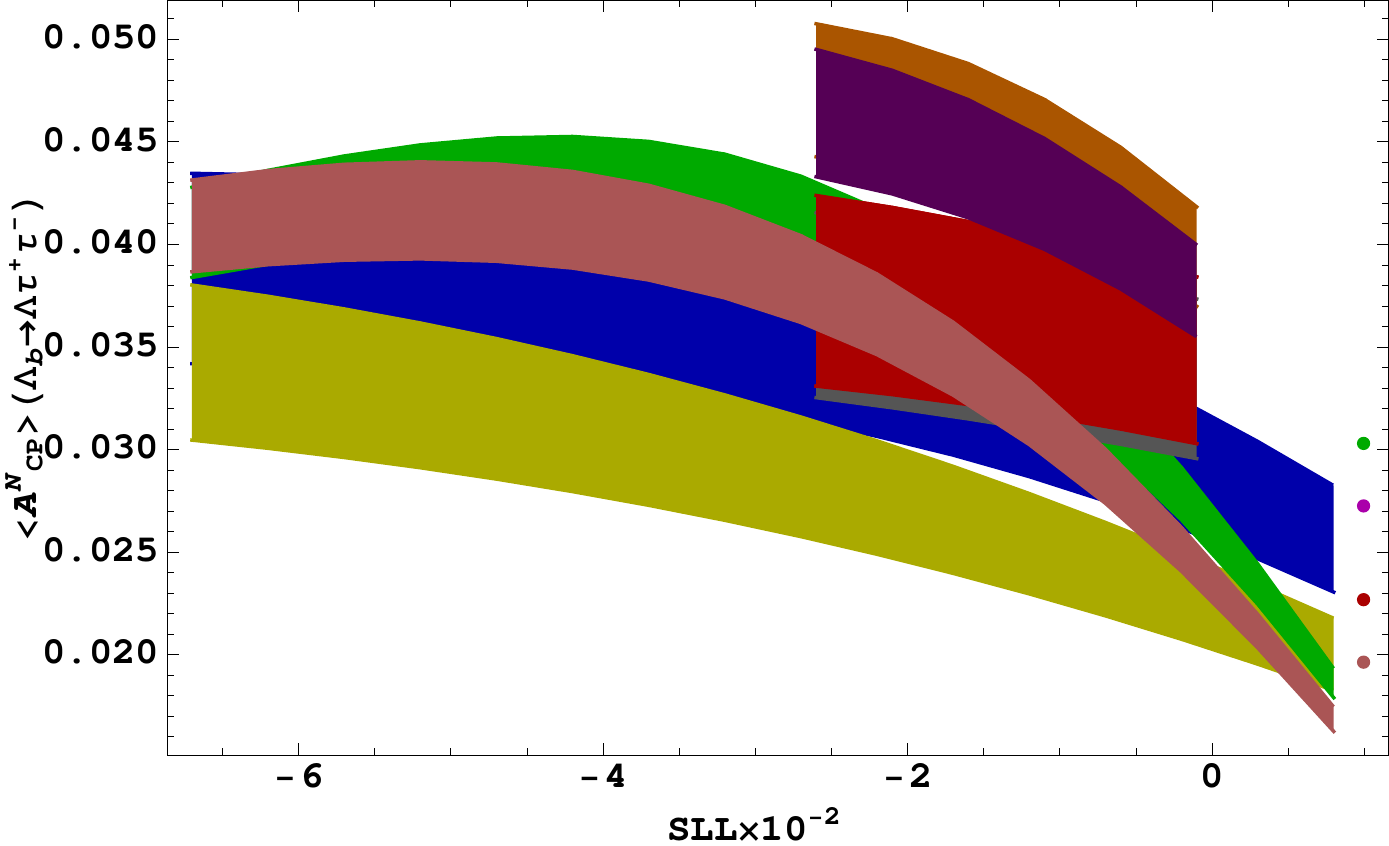}  \put (-100,150){(b)}%
\end{tabular}%
\caption{Normal polarized $CP$ violation asymmetry $\mathcal{A}^{N}_{CP}$ as
function of $S_{LL}$ for  $\Lambda_{b}\to \Lambda \mu^+ \mu^- (\tau^+\tau^-)$ for scenarios $\mathcal{S}1$,
$\mathcal{S}2$ and $\mathcal{S}3$. The color and band description is same as in Fig. \ref{UnCPmu}.} \label{NoCPtau}

\end{figure}
\begin{figure}[tbp]
\begin{tabular}{cc}
\includegraphics[scale=0.55]{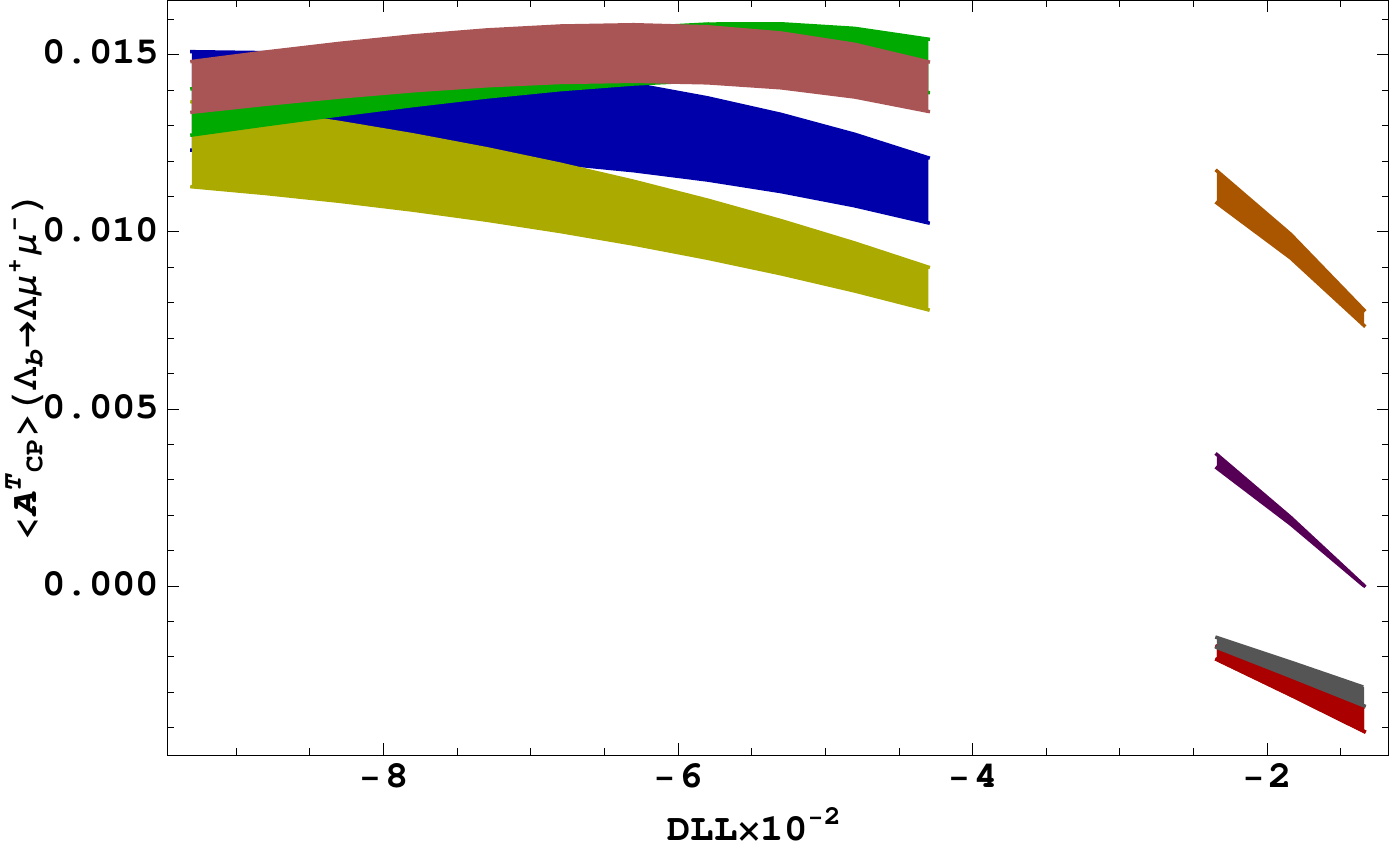}  \put (-100,150){(a)} & %
\includegraphics[scale=0.55]{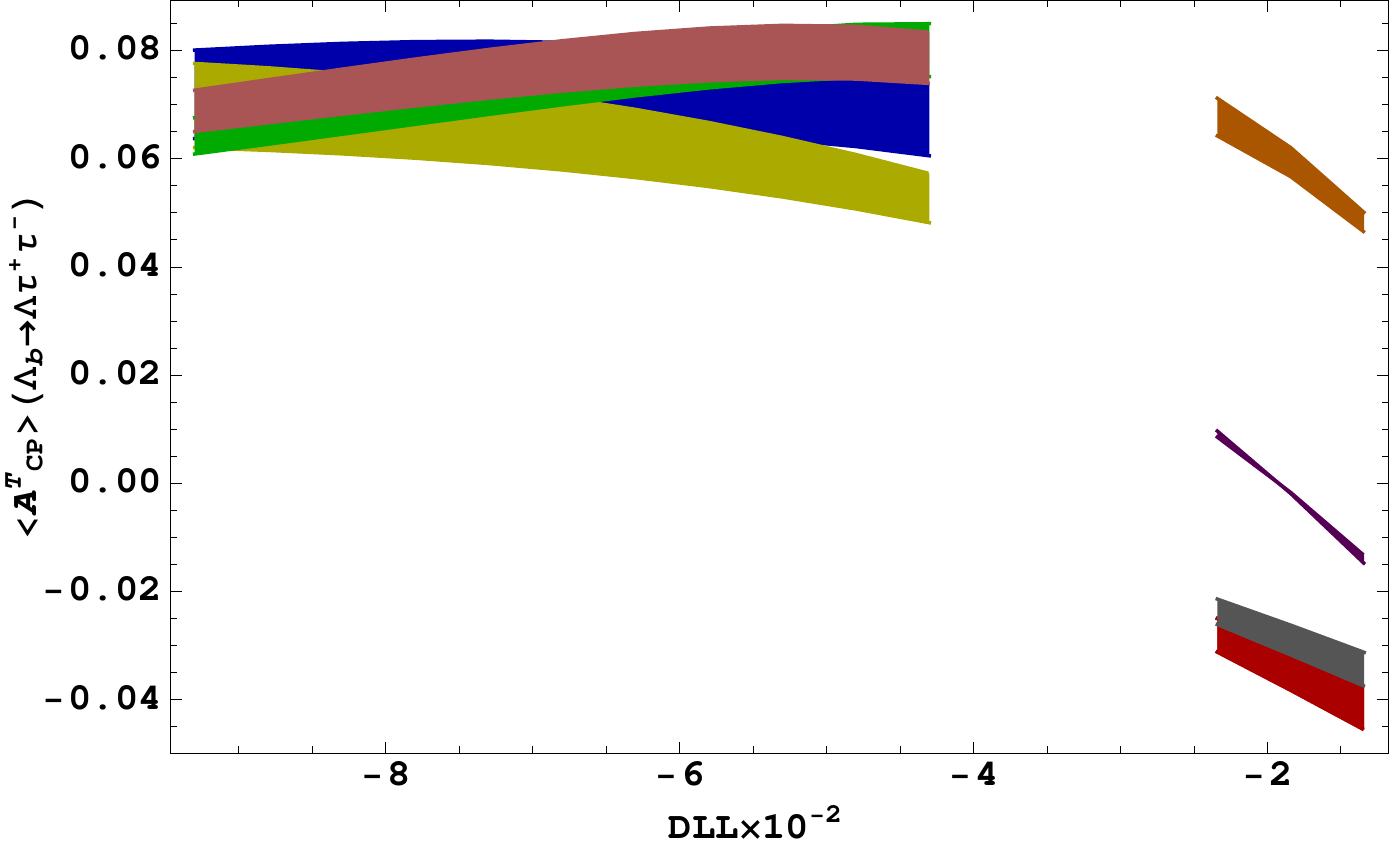}  \put (-100,150){(b)}%
\end{tabular}%
\caption{Transversally polarized $CP$ violation asymmetry $\mathcal{A}^{T}_{CP}$ as
function of $D_{LL}$ for $\Lambda_{b}\to \Lambda \mu^+ \mu^- (\tau^+\tau^-)$ for scenarios $\mathcal{S}1$,
$\mathcal{S}2$ and $\mathcal{S}3$. The color and band description is same as in Fig. \ref{UnCPmu}.} \label{TrCPmu}
\end{figure}
\begin{figure}[tbp]
\begin{tabular}{cc}
\includegraphics[scale=0.55]{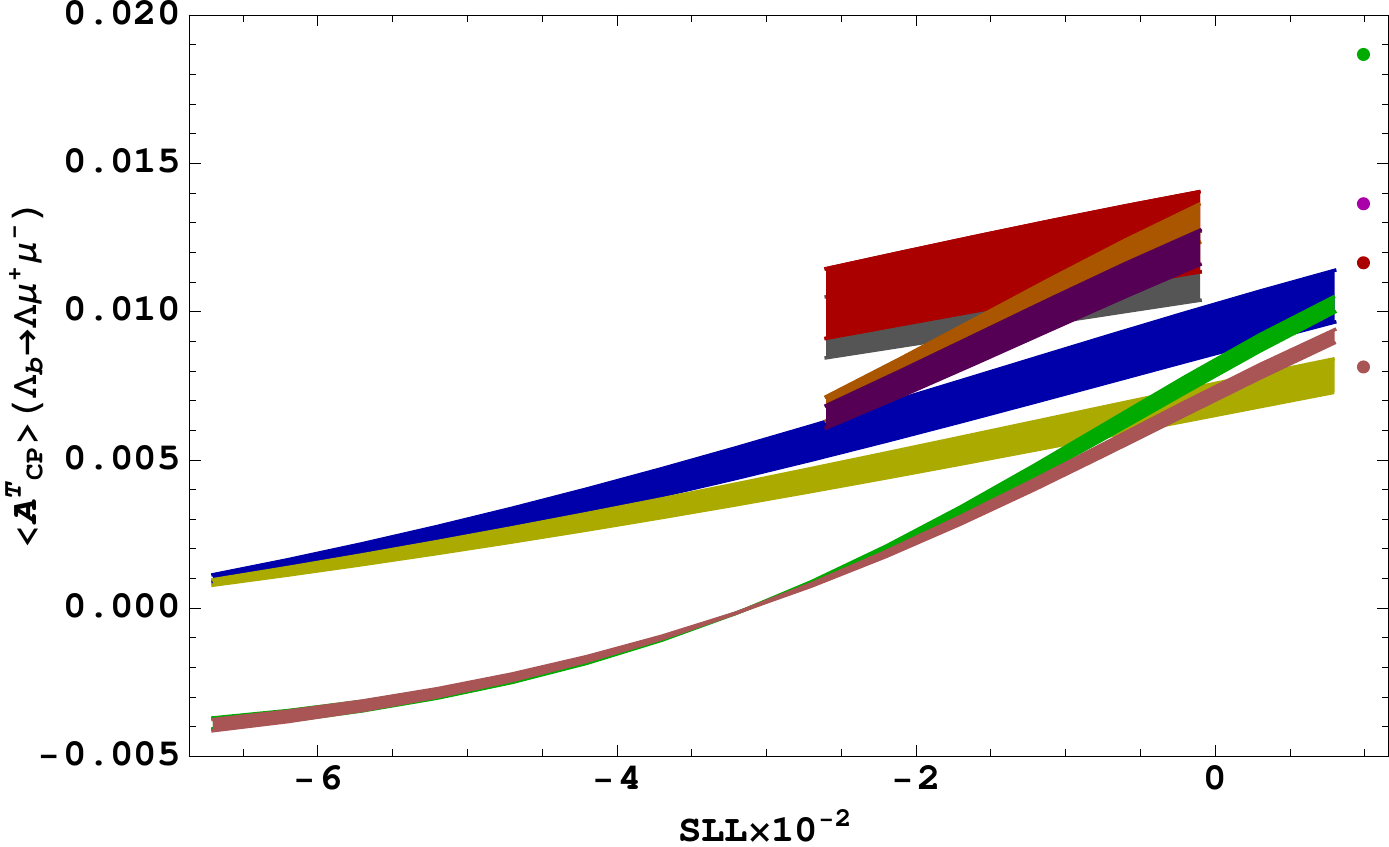}  \put (-100,150){(a)} & %
\includegraphics[scale=0.55]{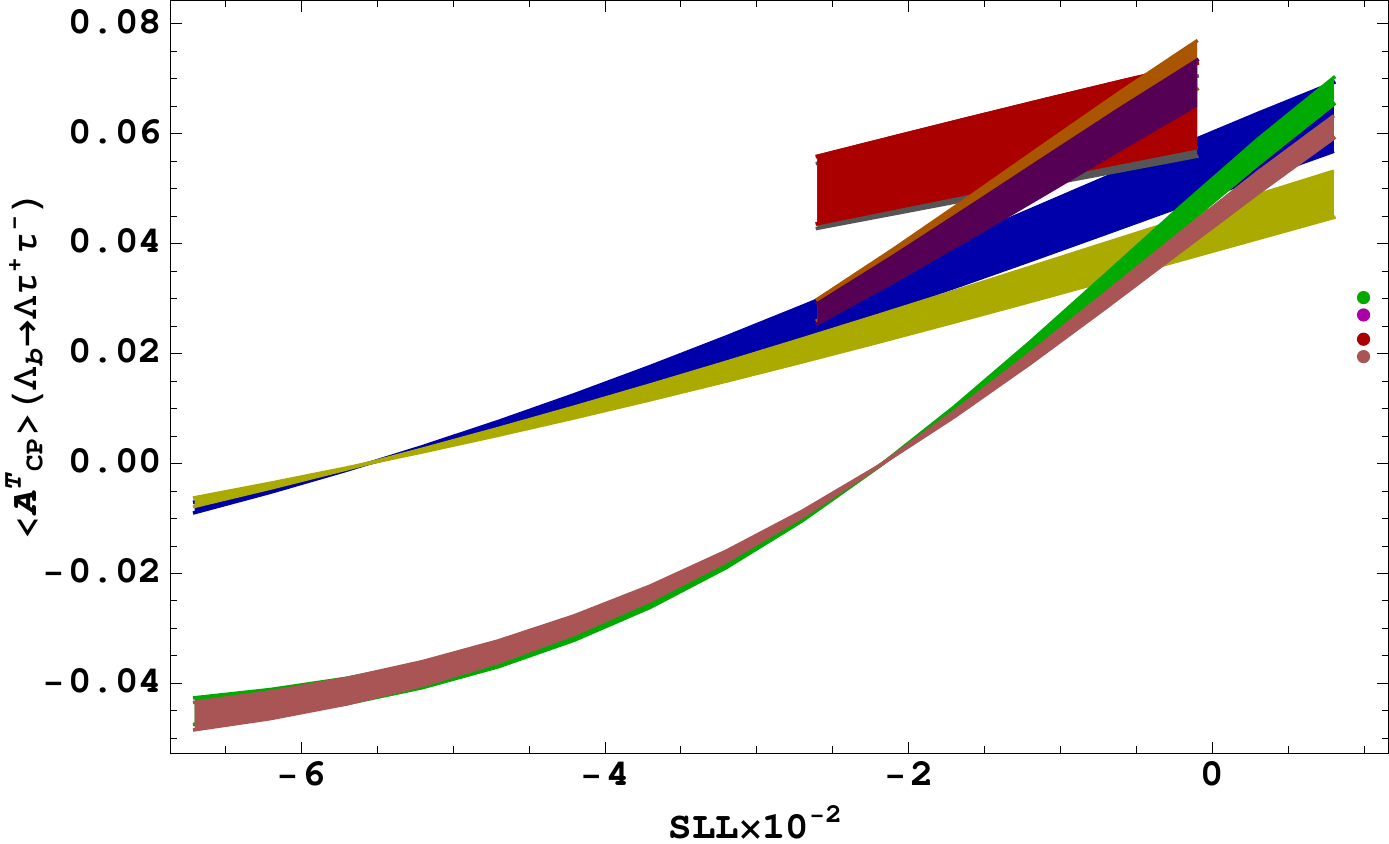}  \put (-100,150){(b)}%
\end{tabular}%
\caption{Transversally polarized $CP$ violation asymmetry $\mathcal{A}^{T}_{CP}$ as
function of $S_{LL}$ for $\Lambda_{b}\to \Lambda \mu^+ \mu^- (\tau^+\tau^-)$ for scenarios $\mathcal{S}1$,
$\mathcal{S}2$ and $\mathcal{S}3$. The color and band description is same as in Fig. \ref{UnCPmu}.} \label{TrCPtau}
\end{figure}

In short, we have analyzed the imprints of NP coming through the neutral $Z^{\prime}$ boson on the unpolarized
and polarized $CP$ violation asymmetries in $\Lambda_{b} \rightarrow \Lambda \ell^{+} \ell^{-}$ decays. In addition motivated by the fact that the $CP$ violation asymmetry is negligible in the SM, we have chosen this observable to explore the effects of $Z^\prime$ in $\Lambda_{b} \rightarrow \Lambda \ell^{+} \ell^{-}$ decays. It has been noticed that the value of unpolarized and polarized $CP$ violation asymmetry is
considerable in both $\Lambda_{b} \rightarrow \Lambda  \mu^{+} \mu^{-}$ and $\Lambda_{b} \rightarrow \Lambda \tau^{+} \tau^{-}$ channels and hence it gives a clear
message of NP arising from the neutral $Z^{\prime}$ boson.
Though the detection of leptons' polarization effects in semileptonic decays is really a daunting task at the experiments such as the ATLAS, CMS and at LHCb, but the fact that these $CP$ violation asymmetries which suffer less from hadronic uncertainties provide a useful
probe to establish the NP coming through the $Z^{\prime}$ model.

\section*{Acknowledgments}

The author M. J. A would like to thank the support by Quaid-i-Azam University through the University Research
Fund. M. A. P.  and I. A. would like to acknowledge the grants (2012/13047-2) and (2013/23177-3) from FAPESP.

\end{document}